\documentclass[twocolumn]{aastex631}

\usepackage{amsmath}
\newcommand\numberthis{\addtocounter{equation}{1}\tag{\theequation}}
\newcommand{\cosmic}{\texttt{COSMIC}}
\newcommand{\legwork}{\texttt{LEGWORK}}
\usepackage{xcolor}
\usepackage{showyourwork}

\newcolumntype{C}[1]{>{\centering\let\newline\\\arraybackslash\hspace{0pt}}m{#1}}

\shorttitle{Applying a metallicity-dependent binary fraction to DWD formation}
\shortauthors{Thiele et al.}

\begin{document}

\title{Applying the metallicity-dependent binary fraction to double white dwarf formation: \\ Implications for LISA}

\correspondingauthor{Sarah Thiele}
\email{sarahgthiele@gmail.com}

\author[0000-0001-7442-6926]{Sarah Thiele}
\affiliation{Department of Physics and Astronomy, University of British Columbia, 6224 Agricultural Road, Vancouver, BC, V6T 1Z1, Canada}
\affiliation{Canadian Institute for Theoretical Astrophysics, University
of Toronto, 60 St. George Street, Toronto, Ontario, M5S 1A7,
Canada}
\author[0000-0001-5228-6598]{Katelyn Breivik}
\affiliation{Center for Computational Astrophysics, Flatiron Institute, 162 Fifth Ave, New York, NY, 10010, USA}
\affiliation{Canadian Institute for Theoretical Astrophysics, University
of Toronto, 60 St. George Street, Toronto, Ontario, M5S 1A7,
Canada}
\author[0000-0003-3939-3297]{Robyn E. Sanderson}
\affiliation{Department of Physics and Astronomy, University of Pennsylvania, 209 South 33rd Street, Philadelphia, PA 19104, USA}
\affiliation{Center for Computational Astrophysics, Flatiron Institute, 162 Fifth Ave, New York, NY, 10010, USA}
\author[0000-0002-0296-3826]{Rodrigo Luger}
\affiliation{Center for Computational Astrophysics, Flatiron Institute, 162 Fifth Ave, New York, NY, 10010, USA}

\begin{abstract}
Short-period double white dwarf (DWD) binaries will be the most prolific source of gravitational waves (GWs) for the Laser Interferometer Space Antenna (LISA). DWDs with GW frequencies below $\sim1\,\rm{mHz}$ will be the dominant contributor to a stochastic foreground caused by overlapping GW signals. Population modeling of Galactic DWDs typically assumes a binary fraction of $50\%$ and a log-uniform Zero Age Main Sequence (ZAMS) orbital period distribution. However, recent observations have shown that the binary fraction of close, solar-type stars exhibits a strong anti-correlation with metallicity which modulates the ZAMS orbital period distribution below $10^4$ days. In this study we perform the first simulation of the Galactic DWD population observable by LISA which incorporates an empirically-derived metallicity-dependent binary fraction, using the binary population synthesis suite \cosmic\ and a metallicity-dependent star formation history. We compare two models: one which assumes a metallicity-dependent binary fraction, and one with a binary fraction of 50\%. We repeat our analysis for three different assumptions for Roche-lobe overflow interactions. We find that while metallicity impacts the evolution and intrinsic properties of our simulated DWD progenitor binaries, the LISA-resolvable populations of the two models remain roughly indistinguishable. However, the size of the total Galactic DWD population orbiting in the LISA frequency band is reduced by more than half when accounting for a metallicity-dependent binary fraction for two of our four variations, which also lowers the effective foreground. The LISA population remains unchanged in number for two variations, highlighting the sensitivity of the population to binary evolution prescriptions.
\end{abstract}


\section{Introduction} \label{sec:intro}
Most stars in the Galaxy will end their lives as white dwarfs. Of the stars which are born with a binary companion, many will undergo interactions which bring the two stars closer together, eventually forming a close double white dwarf (DWD). Close DWDs, with orbital periods shorter than $\sim2.75\,\rm{hr}$ are the largest source by number of mHz gravitational waves (GWs) in the Galaxy \citep[e.g.,][]{LISAMissionProposal}. The Laser Interferometer Space Antenna (LISA) is expected to resolve at least $10^4$ individual DWD binaries in the Milky Way and will also observe GW emission from the entire Galactic DWD population through the unresolved foreground created by overlapping signals at sub-mHz frequencies \citep[e.g.,][]{Nelemans2001a, Ruiter2010, Nissanke2012, Yu2013, Korol2017, Lamberts2019, Breivik2020a}. The resolved population will enable the study of several important aspects of binary evolution like the strength of tides \citep{Valsecchi2012}, the stability of mass transfer in DWD systems \citep[e.g.,][]{Marsh2004, Shen2015, Gokhale2007, Sepinsky2014, Kremer2015}, and the separation distribution of close DWDs \citep{Korol2021} as well as provide a probe of Galactic structure \citep{Korol2019} and the Local Group \citep{Korol2018}. The shape and strength of the Galactic DWD foreground can also be used as a tool to study the structure of the Milky Way \citep{Benacquista2006, Breivik2020b}.

The transition between individually resolved DWDs and the confusion limited, or unresolved, DWD foreground is expected to occur near $\sim1\,\rm{mHz}$ frequencies \citep[e.g. ][]{Ruiter2010}. In the confusion limited regime, more than one binary radiates GWs in each LISA frequency bin, thus creating a superposition of signals which are unable to be disentangled. There are multiple parameters which shape the Galactic DWD foreground. Assuming the evolution of each DWD system is driven solely by GW emission, the frequency derivative scales proportionally to $f_{\rm{GW}}^{11/3}$, causing a pileup of DWDs at lower frequencies \citep[e.g.][]{Breivik2020b}. The spatial density of DWDs in the Galaxy which defines the distance to each binary also impacts the foreground amplitude, since GW strain scales inversely with distance. Star formation history assumptions combine these two effects by assigning ages and positions to each DWD in the population which determine the present day orbital (and thus GW) frequencies as well as distances to each source.

When viewed strictly as a source of noise, the unresolved Galactic DWD foreground is the dominant noise source for LISA in the sub-mHz part of LISA's frequency range. This extra noise above the detector noise floor affects the detection of all other LISA sources including extreme mass ratio inspirals \citep[e.g.,][]{Berti2006, Barack2007, Babak2017, Moore2017}, merging black holes with masses between $10^4$--$10^7\,M_{\odot}$ \citep[e.g.,][]{Klein2016, Bellovary2019}, and cosmological GW backgrounds \citep[e.g.,][]{Bartolo2016, Caprini2016, Caldwell2019}. For sources which have signals buried by the Galactic DWD population, the foreground must be carefully analyzed and subtracted \citep[][]{Adams2014,Cornish2020,Littenberg2020,Boileau2021}. The number of resolved DWDs and the height of the unresolved DWD foreground are a direct consequence of the number of DWD progenitors which form and evolve over the Milky Way's history. 

While the binary fraction remains approximately constant across a large metallicity range ($-1.5 \leq$ [Fe/H] $< 0.5$) for wide binaries, close OB stars, and the stellar Initial Mass Function (IMF) \citep{Moe2017, Moe2019}, the binary fraction for solar-type star systems with orbital period $P_{\rm{orb}}\leq 10^4$ days (separation $a \leq 10$ AU) shows a strong anti-correlation with metallicity \citep[e.g.][]{Badenes2018, Moe2019, Mazzola2020, Price-Whelan2020}. Because close DWDs are the remnants of close, solar-type binary stars, this anti-correlation plays an important role in the formation, evolution, and characteristics of the DWD population that LISA will observe.

To date, population synthesis studies of the Galactic population of close DWDs have either assumed a $100\%$ binary fraction or a $50\%$ binary fraction, such that for every three stars formed, two reside in a binary system \citep{Nelemans2001a, Yu2013, Korol2017, Lamberts2019}. In this study, we investigate the effects of a metallicity-dependent binary fraction on the formation and evolution of DWDs. To this end, we create synthetic present-day Milky Way-like galaxies of DWDs and specifically select systems with GW signals that may be observable by the space-based detector LISA. Throughout, we make comparisons between the standard assumption of a constant $50\%$ initial binary fraction (hereafter model F50) and one with a metallicity-dependent binary fraction (hereafter model FZ).  

In Section \ref{sec:simulations} we discuss our assumptions used in simulating DWD populations and detail the process to produce present-day synthetic Milky-Way-like galaxies. In Section \ref{sec:LISA_obs} we review the derivation of LISA detectability for circular DWD populations at mHz frequencies. In Section \ref{sec:results} we detail results showing how a metallicity-dependent binary fraction affects the formation and evolution of DWD populations assuming four sets of of binary evolution assumptions. In Section \ref{sec:LISA_met} we detail the metallicity dependence of the LISA DWD foreground and resolved population. Finally, we consider how a metallicity-dependent binary fraction impacts the height of the Galactic DWD foreground in LISA for different binary evolution assumptions in Section \ref{sec:model_compare} and conclude in Section \ref{sec:conclusions}.  

\section{Simulating a Galactic DWD population}\label{sec:simulations}
In this section we describe the setup of our DWD simulations using the binary population synthesis suite \cosmic, and the process to scale these simulations to create Milky Way-like galaxies using the star formation history of galaxy \textbf{m12i} in the Latte suite of the FIRE-2 simulations \citep{Wetzel2016, Hopkins2018} and stellar positions assigned according to the the Ananke framework \citep{Sanderson2020}. 

\subsection{Metallicity-dependent binary fraction}\label{subsec:metbinfrac}
The close binary fraction in the Galaxy has been empirically shown to depend on metallicity. This is manifested through an explicit dependence of the companion frequency on orbital period \citep{Moe2021}. When anchored to a log-normal orbital period distribution, the \emph{frequency} of solar-type binary companions is skewed to shorter orbital periods. The companion frequency differs from the binary fraction: the companion frequency is the probability of a star having a companion at a given orbital period, and the integration of this probability over all periods results in the binary fraction of the population. The metallicity dependence of the companion frequency is strongly identified only below $P_{\rm{orb}}=10^4$ days, whereas for $P_{\rm{orb}}>10^6$ it appears to be metallicity-invariant \citep{Moe2019}. We follow \citet{Moe2019}'s choice to linearly interpolate between the two regimes due to a dearth of observational data in this range.

\begin{figure}
	\includegraphics[width=\columnwidth]{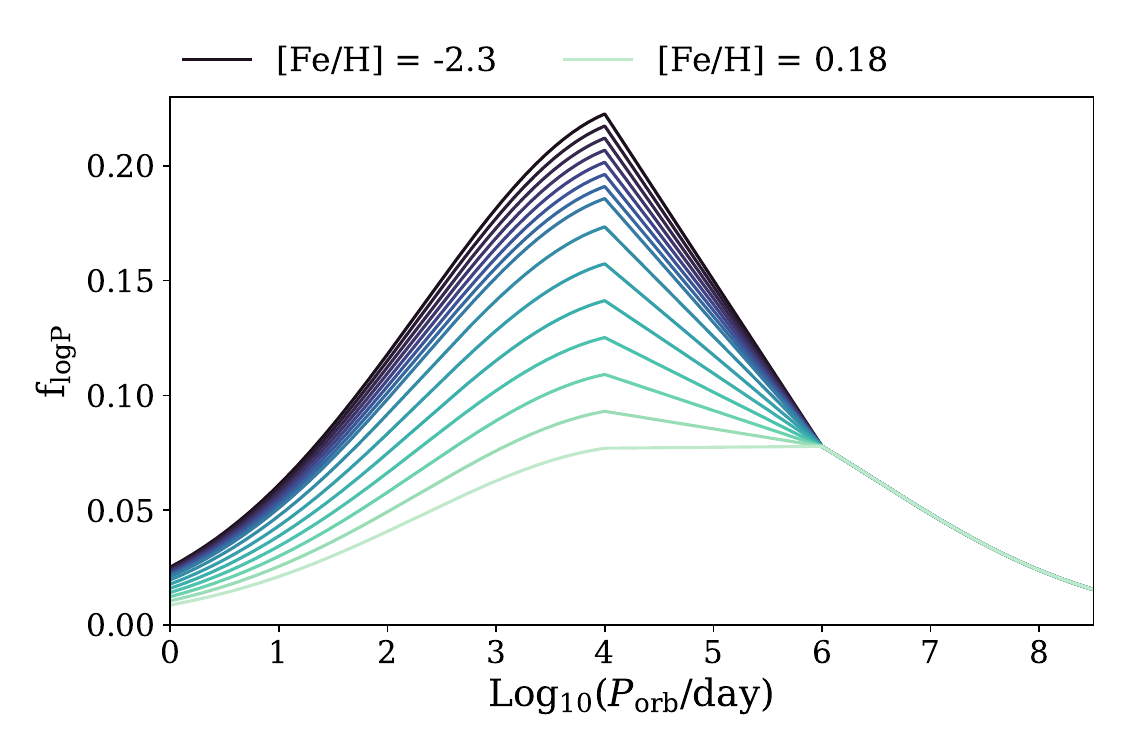}
    \caption{The distribution of the companion frequency f$_{\rm{logP}}$ for solar-type stars is dependent on both ZAMS orbital period and metallicity. Each curve is coloured by metallicity, with the darkest curve corresponding to the lowest-metallicity bin in our simulations at [Fe/H] $=-2.3$ and the lightest curve corresponding to our highest-metallicity bin at [Fe/H] $=0.18$. Above $\log_{10}(P_{\rm{orb}}/\rm{day})\sim 4$, the companion frequency distribution converges to the same curve for all metallicity bins as no identified metallicity dependence at those periods exists in current literature.}
    \label{fig:met_dep_porb}
    \script{met_dep_porb.py}
\end{figure}

Figure \ref{fig:met_dep_porb} shows the companion frequency f$_{\rm{logP}}$ for solar-type stars as a function of the logarithm of orbital period, mirroring the inner binaries curve of Figure 2 in \citet{Moe2021} and scaling to an overall multiplicity frequency of 0.67 (equivalent to an integrated binary fraction of 50\%) at solar metallicity. To obtain the normalization of the companion frequency for close binaries ($P_{\rm{orb}}<10^4$), we fit the results presented in \citet{Moe2019} using linear regression to obtain a piecewise relation between the metallicity, [Fe/H] and close binary fraction $f_{\rm{b}}$ as

\[f_{\rm{b}}= \begin{cases} 
      - 0.0648 \, \rm{ [Fe/H]} + 0.3356, & [\rm{Fe/H}]\leq -1.0 \numberthis \\
     -0.1977 \, \rm{ [Fe/H]} + 0.2025, & [\rm{Fe/H}]> -1.0  , 
   \end{cases}
\]

\noindent where we convert between [Fe/H] and metallicity Z, assuming all stars have solar abundance such that
\begin{equation}
\label{eq:metallicity}
    [\rm{Fe/H}] = \log_{10}\Big(\frac{\textit{Z}}{\textit{Z}_{\odot}}\Big),
\end{equation}
\noindent where we assume $Z_{\odot}=0.02$.

We count all higher-order companions as secondary, or binary, components. We show curves for all 15 of our metallicity bins, increasing in metallicity from darkest to lightest curve colour. The lowest metallicity ($\rm{[Fe/H]}=-2.3$) shows the highest propensity for short-period binaries, in contrast to higher metallicity systems which have a frequency distribution with a relatively constant companion frequency for orbital periods between $10^4$ and $10^6$ days.

While we simulate binaries across the entire orbital period distribution, the close binary fraction, which exhibits the discussed metallicity dependence, will be that which impacts the LISA population since systems with $\log_{10}(P_{\rm{orb}}/$day)$>4$ will be unlikely to evolve enough to enter the LISA band by present day.

\subsection{Binary population models}
\label{sec:bin_pop}
We simulate the evolution of DWD progenitor populations using \cosmic\footnote{https://cosmic-popsynth.github.io}, an open-source Python-based rapid binary population synthesis suite which employs single and binary star evolution using \texttt{SSE/BSE} \citep{Hurley2000, Hurley2002}. Several modifications have been added to \cosmic\ which incorporate updates for massive star evolution and binary interactions. For a detailed description of these modifications see \citet{Breivik2020a}. \cosmic\ has been used in several studies to examine the effects of binary evolution on binary populations from blue stragglers \citep{Leiner2021} and heartbeat stars \citep{Jayasinghe2021}, to white dwarf populations \citep{Kremer2017,Breivik2018,Kilic2021}, to merging compact object populations in isolated binaries \citep{Zevin2020b, Zevin2020a, Zevin2021, Wong2021, Mandhai2021} and in dynamical environments around super-massive black holes \citep{Stephan2019, Wang2021}. 

\cosmic\ is especially useful for efficient generation of large populations of compact binaries. Instead of choosing a fixed number of binary stars for each simulation, \cosmic\ iteratively simulates populations until parameter distributions of the binary population converge to a stable shape as more binaries are added. This process is quantified through the $match$ parameter inspired by matched filtering techniques \citep[e.g. Eq. 6 of ][]{Chatziioannou2017} defined as

\begin{equation}
   match = \frac{\sum_{k=1}^{N}P_{k,i}P_{k,i+1}}{\sqrt{\sum_{k=1}^{N}P_{k,i}P_{k,i}\sum_{k=1}^{N}P_{k,i+1}P_{k,i+1}}},
\end{equation}
where $P_{k,i}$ represents the height of bin $k$ on the $i\rm{th}$ iteration \citep{Breivik2020a}. In this study, we simulate binaries until $\log_{10}(1-match) \leq -5$ for the masses and orbital periods of each DWD population at the formation of the second WD. Since all DWD progenitor binaries simulated with \cosmic\ are circularized through mass transfer or tides before the second WD forms \citep[e.g.][]{Marsh2004, Gokhale2007, Sepinsky2014, Kremer2015}, we do not consider convergence of DWD eccentricities.

The masses and orbital periods at the formation of the second-formed WD span a wide range depending on the WD binary component types, thus we consider four DWD combinations: two helium WDs (He + He), a carbon-oxygen WD orbiting a helium WD (CO + He), two carbon-oxygen WDs (CO + CO), and an oxygen-neon orbiting a helium, carbon-oxygen, or oxygen-neon WD (ONe + X). For each DWD type we simulate a grid of $15$ metallicities spaced uniformly in $\log_{10}(Z)$ between $Z=10^{-4}$ to $0.03$, to account for the limits of the \citet{Hurley2000} stellar evolution tracks employed in \cosmic. This results in a total of $60$ populations across all DWD types and metallicities for each set of model assumptions. The output of \cosmic\ contains information limited to intrinsic binary properties like mass and orbital period. External parameters like Galactic position and orientation are assigned in a post-processing scheme which uses metallicity-dependent positions and ages from the Ananke framework of galaxy \textbf{m12i} from the Latte Suite of the FIRE-2 simulations (see Section~\ref{sec:FIRE} for details).

We assume that the Zero Age Main Sequence (ZAMS) masses, orbital periods and eccentricities for each binary are independently distributed. We choose primary masses following \citet{Kroupa2001}, a flat mass ratio distribution \citep{Mazeh1992, Goldberg1994} and a uniform eccentricity distribution following \cite{Geller2019}. For our simulations which follow a metallicity-dependent binary fraction (model FZ), a skewed log-normal orbital period distribution is sampled in accordance with the metallicity-dependent companion frequency of \citet{Moe2021} as discussed in Section \ref{subsec:metbinfrac}. For our simulations which assume a constant 50\% binary fraction (model F50), we use a log-uniform period distribution following Opik's Law similar to previous studies \citep[e.g.][]{Nelemans2001a, Toonen2012, Korol2017, Lamberts2019}. We assume a $100\%$ binary fraction in the F50 COSMIC simulations to reduce computation time and scale the simulations to a constant $50\%$ binary fraction in a post-processing scheme. We initialize all binaries with the same evolution time of $13.7\,\rm{Gyr}$ to capture all potential evolution within a Hubble time. Although we simulate binaries across the full orbital period distribution, we discard DWD binaries which form with separations $a\geq 1000\,R_\odot$ during post-processing, since these are unlikely to evolve into the LISA band by present day.

We consider a fiducial set of assumptions which follow the \cosmic\ defaults described in \citet{Breivik2020a} except for the treatment of Roche-lobe overflow (RLO). The stability of RLO mass transfer is determined using critical mass ratios resulting from radius-mass exponents \citep{Webbink1985, Hurley2002}, where the critical mass ratio is defined as the ratio of the donor to accretor mass. We assume critical mass ratios following \citet{Claeys2014} which reduce the standard critical mass ratio assumptions from \citet{Hurley2002} for main sequence (MS) donors by $\sim50\%$ from $3$ to $1.6$ based on the models of \citet{deMink2007} and treat WD accretors separately following the models of \citet{Soberman1997}. We increase the mass loss rate from the donor following Equation 11 of \citet{Claeys2014}. The amount of mass lost during RLO from the donor is limited by the overflow factor of the donor radius to its Roche radius following \citet{Hurley2002}. The amount of mass accepted by the accretor is limited to $10$ times the accretor's mass divided by the accretor's thermal timescale. Finally, for RLO mass loss which becomes unstable and leads to common envelope (CE) evolution we assume that the donor's binding energy is calculated according to the fits detailed in Appendix B of \citet{Claeys2014} and that orbital energy is deposited with $100\%$ efficiency into unbinding the common envelope ($\alpha=1$).

The LISA-detectable DWD populations that result from our simulations may vary depending on the assumptions made regarding binary interactions. In order to explore the range of results, we complete this study for three binary evolution parameter variations on top of our fiducial set of assumptions. For each variation, we consider models FZ and F50 as done in the fiducial case described above. In variation $q3$, we vary the assumption for the critical mass ratios at which a RLO interaction remains stable or becomes unstable from our fiducial assumptions. The critical mass ratio $q_c$ is increased to $3.0$ and thus allows stable mass transfer for more massive RLO donors. In variations $\alpha25$ and $\alpha5$, we modify the common envelope ejection efficiency to be either much less ($\alpha=0.25$) or more ($\alpha=5$) than in our fiducial assumption ($\alpha=1$) to capture the range of possible CE ejection efficiencies quoted in the literature  \citep[e.g.][]{Zorotovic2010, Fragos2019}. Larger common envelope ejection efficiencies lead to wider post-CE separations, while smaller ejection efficiencies either lead to closer post-CE separations or stellar mergers where the envelope ejection fails. Between all variations, all other prescriptions for binary evolution remain identical besides the one varied parameter. A summary of all models and variations explored in this work is given in Table \ref{tbl:models}. We delve in depth into the fiducial results in each section, and then give a brief overview of how the results change for each parameter variation. 

\begin{table}[]
    \centering
    \begin{tabular}{|c|c|c|}
       \hline
       model & binary fraction & orbital period distribution \\
       \hline
       FZ & \citet{Moe2019} & \citet{Moe2021}\\
       F50 & 50\% & flat in $\log(P_{\rm{orb}}/\rm{day})$\\
       \hline
       \hline       
       variation & parameter change & binary evolution change \\
       \hline
       fiducial  & none & none \\
       $\alpha25$ & $\alpha=0.25$ & reduced CE efficiency \\
       $\alpha5$  & $\alpha=5$ & increased CE efficiency\\
       $q3$ & $q_c=3$ & increased critical mass ratio \\
       \hline
    \end{tabular}
    \caption{Summary of the models and binary evolution assumption variation simulations that are explored in this study.}
    \label{tbl:models}
\end{table}

\subsection{A metallicity-dependent SFH: Convolving with the FIRE-2 models}
\label{sec:FIRE}
To create Milky-Way-like galaxies which integrate the metallicity-dependent binary fraction, we use the metallicity-dependent ages and positions of galaxy \textbf{m12i} from the ``Latte" suite of the FIRE-2 simulations \citep{Hopkins2015, Wetzel2016, Hopkins2018} to create synthetic, Milky-Way-like DWD populations. 

The \textbf{m12i} galaxy provides particle mass resolution of 7070 $M_\odot$ per star particle. Each star particle has an associated metallicity, position, and age, which is combined with the output of \cosmic\ to assign DWDs to each star particle by matching its metallicity to our \cosmic\ metallicity grid. The positions of each DWD are assigned using the Ananke framework since multiple DWD binaries can form within a single star particle. Specifically, we use an epanechnikov kernel where the kernel size is inversely proportional to the local density to assign the radial component of spherically symmetric offsets from the center of each star particle following \citet{Sanderson2020}. 

\begin{figure}
	\includegraphics[width=\columnwidth]{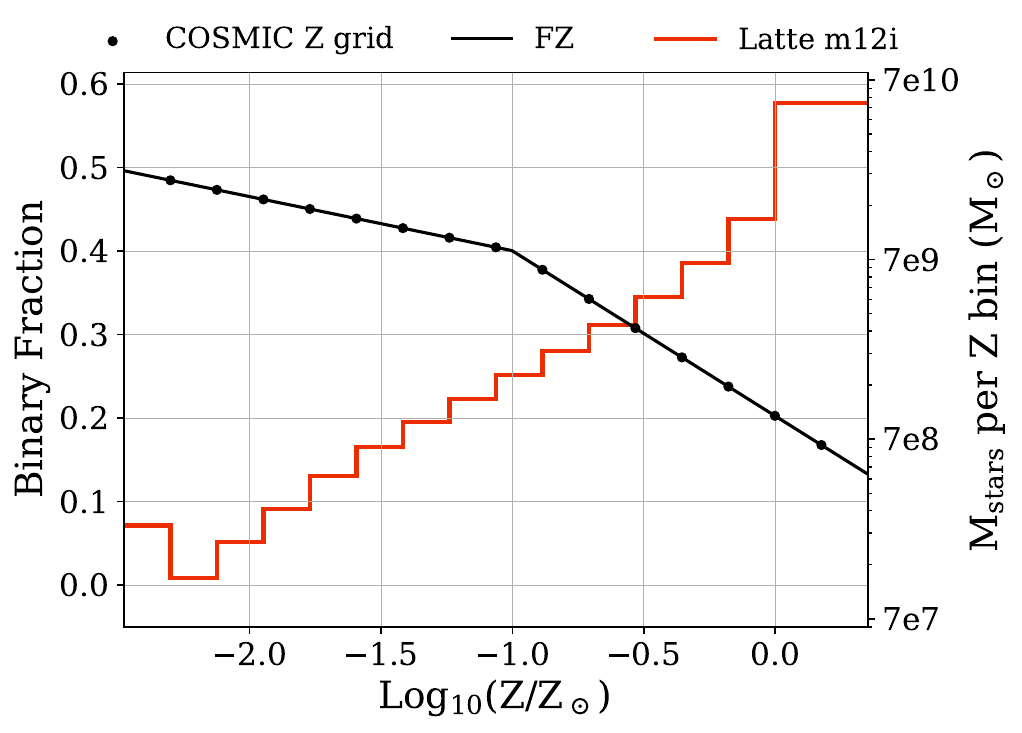}
    \caption{The metallicity-dependent binary fraction for close solar-type binaries with $P_{\rm{orb}}< 10^4$ days (black) plotted against the logarithm of $Z/Z_\odot$, with scatter points denoting the location of the metallicity grid used in our \cosmic\ simulations. The secondary axis shows the amount of mass in stars formed within each metallicity bin from galaxy \textbf{m12i} of the ``Latte" suite in the FIRE-2 simulations as a red histogram. The amount of stellar mass formed with super-solar metallicities dominates the distribution. Note that the primary axis shows $f_{\rm{b}}(Z)$ in linear scale, and the secondary axis shows amount of mass formed in log-scale. The opposing trends of these two distributions compete throughout this study.}
    \label{fig:sfh_vs_fb}
    \script{sfh_vs_fb.py}
\end{figure}

The metallicity-dependent close binary fraction for solar-type binaries ($P_{\rm{orb}} < 10^4$ days), or simply the close binary fraction from here on, is shown in black in Figure \ref{fig:sfh_vs_fb} along with the mass in star particles from galaxy \textbf{m12i}, shown in red, as a function of our metallicity grid. The close binary fraction, $f_{\rm{b}}(Z)$, drops drastically across metallicity while the mass formed in \textbf{m12i} increases significantly. These two opposing trends compete throughout this study along with the impact of metallicity on single star evolution to form the final numerical distribution of systems in our DWD populations. 

Since our \cosmic\ simulations assume a binary fraction of $f_{\rm{b}} = 1$ for model F50, we scale the amount of mass sampled at ZAMS required to produce our \cosmic-generated population of DWDs ($M_{\text{b,ZAMS}}$) to the proper amount of mass sampled in single and binary stars ($M_{\text{ZAMS,sim}}$). We do this by sampling single stars and primary masses of binary stars from the \cite{Kroupa2001} IMF and sampling secondary masses of the binary stars from a uniform mass distribution, where the number of binaries is calculated to ensure $f_{\rm{b}}=0.5$. From this sample, we obtain the ratio of mass in single stars to the mass in binary stars, $R(f_{\rm{b}})$. For model F50, the ratio is a constant $R(f_{\rm{b}}) = 0.64$. The total amount of ZAMS mass in single and binary stars is then $M_{\text{ZAMS,sim}} =  M_{\text{b,ZAMS}} (1 + R(f_{\rm{b}}))$. For model FZ, no scaling is applied since our \cosmic\ simulations already have the population-wide binary fraction incorporated, and thus $M_{\text{ZAMS,sim}} =  M_{\text{b,ZAMS}}$.

Once we determine the total ZAMS mass required to produce our simulated population for a given metallicity, the number of DWDs formed per unit solar mass at metallicity $Z_i$ is
\begin{equation}
    n_{\rm{DWD}}(Z_i) = \frac{N_{\rm{DWD, sim}}(Z_i)}{M_{\rm{ZAMS, sim}}(Z_i)}.
\end{equation}

\noindent The number of DWD’s per \textbf{m12i} star particle at metallicity $Z_i$ is then
\begin{equation}
    N_{\rm{DWD},\star} (Z_i) = n_{\rm{DWD}}(Z_i)\, M_\star,
\end{equation}

\noindent where $M_\star = 7070\,M_\odot$ is the mass per \textbf{m12i} star particle. Since $N_{\rm{DWD},\star} (Z_i)$ is not an integer, we treat the decimal component as the probability that the star particle contains an extra DWD in addition to the integer number. For each star particle, we sample with replacement $N_{\rm{DWD},\star}(Z_i)$ DWDs from the corresponding simulated \cosmic\ population at that metallicity and assign the ZAMS birth time of each DWD to the formation time of the star particle. For most DWD types there is more than one DWD binary system assigned to each \textbf{m12i} star particle.

If the DWD formation time is less than the age of the star particle, we evolve the DWD over the remaining time between its formation and star particle age, $t_{\rm{evol}}$, to produce the present-day population. Once a DWD is formed, we assume that the binary evolves only due to the emission of GWs. Due to tidal effects and mass transfer between their progenitor binaries, all DWDs in our simulations are circular, thus eccentricity does not need to be considered. The orbital evolution over the time $t_{\rm{evol}}$ is then simply defined according to \citet{Peters1964} as
\begin{equation}
    a_f = (a_i - 4\beta t_\text{evol})^{1/4},
\end{equation}
where $a_i$ and $a_f$ are the DWD separations at formation and present day respectively, and 
\begin{equation}
    \beta = \frac{64G^3}{5c^5} M_1M_2(M_1+M_2)
\end{equation}
is constant throughout DWD evolution \citep{Peters1964}.

We discard any DWDs for which the sum of their ZAMS birth time, given by the star particle formation time, and DWD formation time is larger than the age of the star particle since the system will not have evolved long enough to become a DWD at present. We further discard any DWDs for which the lower-mass WD overflows it's Roche lobe before present day, because the outcomes of these interactions are highly uncertain and their treatment is outside the scope of this work \citep[e.g., ][]{Shen2015, Kremer2017}. This choice does not significantly affect our results since stably accreting WD binaries are unlikely to contribute appreciably to the Galactic DWD foreground due to their small mass ratios at frequencies below 1 mHz \citep{Breivik2018}. The separation at which the lower-mass WD overflows its Roche Lobe is defined as

\begin{equation}
    a_{\text{RLO,}\ell} = R_{\ell} \frac{0.6 q_{\ell}^{2/3} + \ln{(1+q_{\ell}^{1/3})}}{0.49 q_{\ell}^{2/3}}
\end{equation}
where $R_{\ell}$ is the radius of the lower-mass WD and $q_{\ell} = M_{\ell}/M_{h}$ is the ratio of the lower- to higher-mass WD components \citep{Eggleton1983}. 
We define the radius of a WD following \citet{Tout1997, Hurley2000} as
\begin{equation}
    R_{\rm{WD}} = \max\Bigg(R_\text{NS}, 0.0115\sqrt{\left(\frac{M_\text{Ch}}{M}\right)^{2/3}-\left(\frac{M}{M_\text{Ch}}\right)^{2/3}} \Bigg) 
\end{equation}
\noindent where $R_\text{NS} = 1.4\cdot 10^{-5}~R_\odot$ is the radius of a neutron star, $M_{\text{Ch}}=1.44 ~M_\odot$ is the Chandrasekhar limit for the mass of a stable WD, and $M$ is the mass of the WD in solar masses. 

For the non-discarded systems, we log the present-day separations from which the present-day orbital frequency $f_\text{orb}$ can be found using Kepler's third law. The GW frequency is then $f_\text{GW} = 2f_\text{orb}$.

\section{LISA detectability}
\label{sec:LISA_obs}
We use \legwork\footnote{https://legwork.readthedocs.io} \citep{Wagg2021} to determine the detectability of our simulated DWD populations for sources with GW frequencies $f_{\rm{GW}}>10^{-4}\,\rm{Hz}$. \legwork\ calculates the position-, orientation-, and angle-averaged signal to noise ratio (SNR) for inspiraling GW sources closely following the derivations of \citet{Flanagan1998} and using the LISA noise power spectral density (PSD) of \citet{Robson2019}. 

To lowest order in the post-Newtonian expansion, the frequency evolution of circular orbits for quadrupole GW emission is defined as
\begin{equation}
    \Dot{f}_n = \frac{48n}{5\pi} \frac{(G\mathcal{M}_c)^{5/3}}{c^5} (2\pi f_\text{orb})^{11/3}.
\end{equation}
We classify DWDs as evolving, or ``chirping", when $\Dot{f}_n \geq 1/T_\text{obs}^2$. For evolving sources, the SNR is

\begin{equation}
   \langle \rho \rangle^2_{\text{circ,evol}} = \int_{f_0}^{f_1}df   \frac{h_{c}^2}{f^2S_n(f)}
\end{equation}

\noindent where $h_c$ is the characteristic strain of the system, $S_n(f)$ is the LISA sensitivity curve of \citet{Robson2019}, and the frequency limits are determined by the orbital evolution over the observation time, $T_{\rm{obs}}=4\,\rm{yr}$. The characteristic strain for circular orbits is 
\begin{equation}
    h_{c}^2 = \frac{2^{2/3}}{3 \pi^{4/3}} \frac{(G \mathcal{M}_c)^{5/3}}{c^3 D_L^2} \frac{1}{f_{\rm orb}^{1/3}},
\end{equation}
\noindent where $\mathcal{M}_c = (M_1M_2)^{3/5}/(M_1 + M_2)^{1/5}$ is the system's chirp mass, and $D_L$ is the systems luminosity distance which we assume to be the distance of each simulated DWD to the Sun. 

For stationary sources, the SNR is modified due to the lack of observable orbital evolution as
\begin{equation}
    \rho_{\text{circ,stat}}^2 = \frac{h_{2}^2T_{\text{obs}}}{ S_n(f_2)}
\end{equation}
with the observation time $T_{\text{obs}}$. Here, $h_2$ is strain amplitude of the source for the second orbital frequency harmonic,
\begin{equation}
   h_2^2 = \frac{2^{22/3}}{5}\frac{(G\mathcal{M}_c)^{10/3}}{c^8D_L^2}(\pi f_\text{orb})^{4/3}
\end{equation}
and is connected to the characteristic strain as
\begin{equation}
    h_2^2 = \frac{\dot{f}_{2}}{f_{\rm{orb}}^2} h_c^2
\end{equation}

The amplitude spectral density for a stationary system is finally defined as $ASD = h_{2} \sqrt{T_\text{obs}}$, such that the SNR for stationary source is simply, $\rho \sim ASD/S_n$.

The Galactic foreground included in the \citet{Robson2019} LISA noise curve was generated using a different binary evolution code and set of model assumptions for DWD formation and evolution \citep{Toonen2012, Korol2017}. Thus, we use the detector curve only and generate an approximate foreground from each of our populations as follows. Instead of performing a full source subtraction algorithm \citep[e.g.][]{Littenberg2020}, which is out of the scope of this work, we calculate the PSD of the Galactic DWD population with a frequency resolution set by the LISA mission time as $1/T_{\rm{obs}}\sim1/4\,\rm{yr}^{-1}\sim8\times10^{-9}\,\rm{Hz}$. We then approximate the foreground as the running median of the PSD with a boxcar window with a width of $10^3$ frequency bins similar to \citet{Benacquista2006}. The Galactic DWD PSD is truncated near $10\,\rm{mHz}$ for both of our models because we remove all DWDs which experience Roche-lobe overflow. In order to smooth the effect of this truncation in our foreground, we fit each running median with fourth-order polynomials for GW frequencies up to $1\,\rm{mHz}$, thus allowing an approximation of the foreground PSD for higher frequencies. These fits are listed in Table~\ref{tbl:fits} where the polynomial is described as
\begin{equation}
\label{eq:fit}
    \log_{10}(\rm{confusion\ fit}/\rm{Hz}) = a\,x^4 + b\,x^3 + c\,x^2 + d\,x + e
\end{equation}
\noindent and $\rm{x}=\log_{10}(f_{\rm{GW}}/\rm{Hz})$. We add the fitted polynomial of the PSD's running median to the LISA noise PSD to obtain a sensitivity curve for each model and variation. 

\begin{table}[]
    \centering
    \begin{tabular}{|c|c|c|c|c|c|}
       \hline
       model & a & b & c & d & e \\
       \hline
       \hline
       fiducial, F50 & -217.8 & -183.9 & -74.5 & -13.7 & -1.0 \\
       \hline
       fiducial, FZ & -268.2 & -243.0 & -100.5 & -18.7 & -1.3 \\
       \hline
       $\alpha25$, F50 & -5520.2 & -6078.3 & -2528.8 & -467.1 & -32.3 \\
       \hline
       $\alpha25$, FZ & -2816.0 & -3066.4 & -1272.7 & -234.7 & -16.2 \\
       \hline
       $\alpha5$, F50 & -178.7 & -142.8 & -58.7 & -11.0 & -0.8 \\
       \hline
       $\alpha5$, FZ & -265.9 & -239.8 & -99.1 & -18.5 & -1.3 \\
       \hline
       q3, F50 & -114.5 & -73.5 & -30.9 & -6.1 & -0.5 \\
       \hline
       q3, FZ & -336.9 & -323.5 & -136.2 & -25.8 & -1.8 \\
       \hline
    \end{tabular}
    \caption{Polynomial fitting coefficients for the confusion foreground fit of Equation~\ref{eq:fit} for each binary fraction model and parameter variation.}
    \label{tbl:fits}
\end{table}

\section{Metallicity effects on the formation and evolution of DWDs}\label{sec:results}

\subsection{DWD types and their formation channels}\label{sec:ini}
As discussed in Section\,\ref{sec:bin_pop}, we consider four DWD sub-types, which each contribute differently to LISA's GW signals: He + He, CO + He, CO + CO, and ONe + X. Each sub-type has a unique distribution in their formation times, initial masses, radii, and orbital periods stemming from variations in their evolution channels and their formation efficiency. Here we describe the general formation scenarios and population properties of Galactic close DWDs which may be observable by LISA.

He WDs are unable to form through single star evolution within the lifetime of the Milky Way. Instead, they originate through interactions in close binary systems or binaries with large eccentricities. Because of this, He WDs are able to form with low component masses on order $\sim 0.1\,M_\odot$, with the majority of He WDs in our simulations having masses between $0.2$--$0.5$ $M_\odot$. He + He DWDs form through evolution of close binary systems, during which their envelopes are both stripped through RLO and CE phase interactions before Helium ignition occurs. The two progenitor stars generally have masses $\lesssim 3\,M_\odot$ which is lower than the progenitors of other DWD types. Our simulated He + He DWDs have an approximately constant distribution of formation times $\gtrsim 2.5\,\rm{Gyr}$. Lastly, since the ZAMS separations are skewed towards shorter values, we also see that the resulting DWD separations are smaller on average than that of other DWD types. 

\begin{figure*}
	\includegraphics[width=\textwidth]{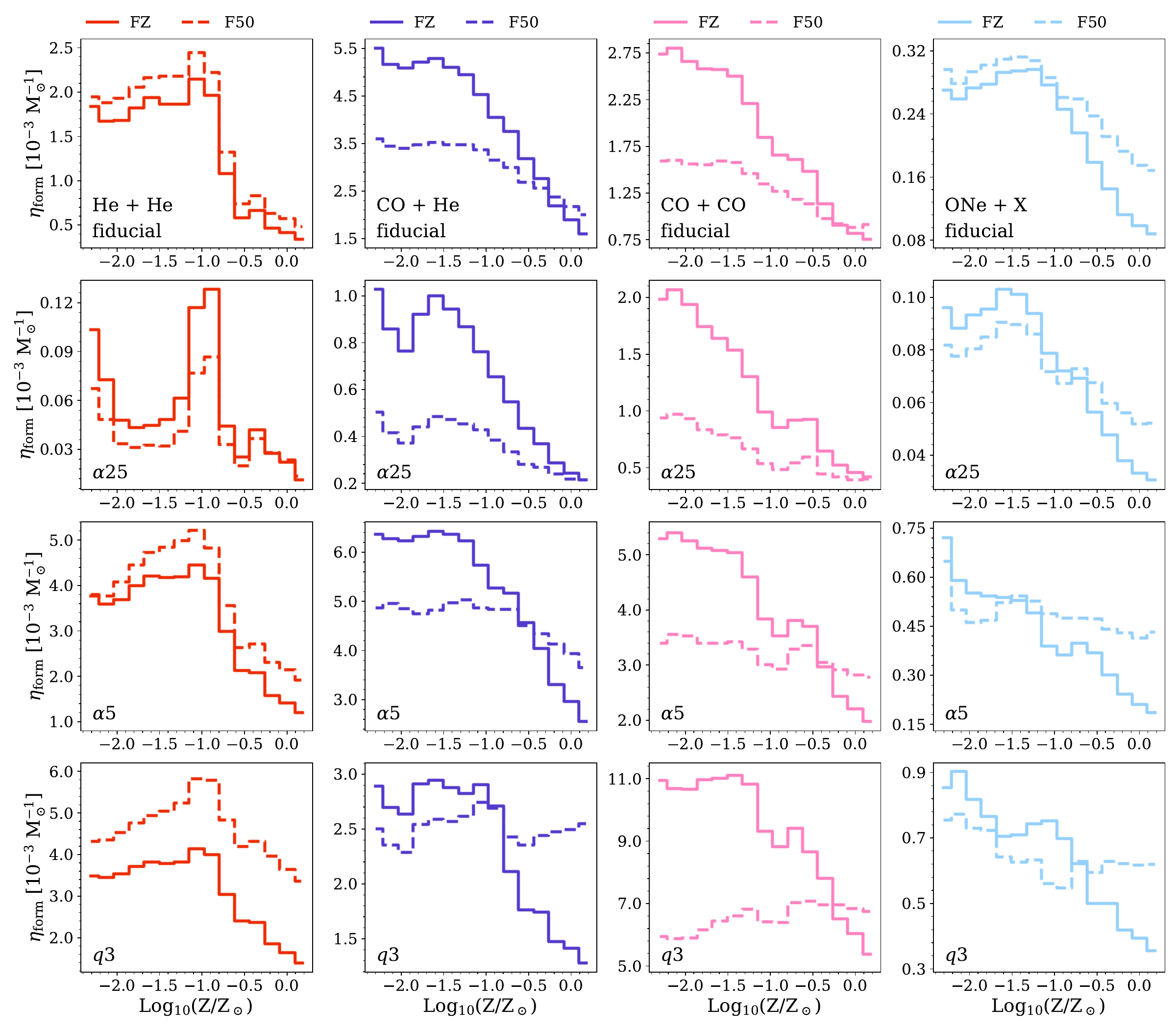}
    \caption{The DWD formation efficiency vs metallicity of DWD populations simulated with \cosmic. Each panel shows the formation efficiency for a given DWD type and variation. The solid lines indicate the formation efficiency for model FZ which incorporates a metallicity-dependent binary fraction. The dashed lines indicate the formation efficiency for model F50, which assumes a constant binary fraction of $50\%$. The DWD formation efficiency drops by a factor of $4$--$5$ for model FZ and a factor of $1$--$5$ for model F50. See Section~\ref{sec:formeff} for a careful description of the trends for each DWD type.}
    \label{fig:form_eff}
    \script{form_eff.py}
\end{figure*}

A CO WD forms when a star is able to begin the helium burning process before its envelope is stripped. Thus to form a CO + He DWD, RLO and CE stages occur after one component experiences core helium burning, but before the other component can. Most close CO + He DWDs form in approximately 2 Gyr after ZAMS and with very short periods because the He WD is formed through the ejection of a common envelope which greatly reduces the orbital separation. Because of these short formation separations, many CO + He DWDs merge before the present day. Due to their asymmetric mass distributions, they have lower chirp masses, but their shorter periods make them important candidates for LISA detection. 

To prevent the two stars' envelopes from being stripped before helium ignition, CO + CO DWDs typically form from progenitors in wider orbits, and the two components may have little to no interaction during their evolution from ZAMS to DWD. CO DWDs thus have a distribution in progenitor separation that extends to larger values than for other DWD types. Most CO + CO DWDs need $> 0.3$ Gyr to form, have component WD masses between $0.35$--$1.0\,M_\odot$, and make up the majority of the DWD population.

ONe WDs are rare and typically form from massive progenitor stars which evolve through the asymptotic giant branch phase, thus resulting in a higher-mass WD. All ONe WDs in our \cosmic\ populations, e.g., have progenitor ZAMS masses above $4\,M_\odot$, and the resulting ONe WDs have a relatively flat distribution of masses from $1.05\,M_\odot$ up to the Chandrasekkhar limit of $1.4\,M_\odot$. Because an ONe WD can have a companion of any other WD type in our study, there is a spread in their distributions for separation, secondary mass, final orbital period, and formation time. In general, however, these systems result from wider separations to allow for the evolution of the ONe component without merging. For example, all initial separations in our \cosmic \ populations have separations $\gtrsim 1.5\,R_\odot$. ONe + X DWDs can form on short timescales, as low as $30\,\rm{Myr}$, for the majority of high-metallicity systems.

\subsection{Metallicity-dependent trends in the formation efficiency of DWDs}\label{sec:formeff}
The number of DWDs formed per unit solar mass of ZAMS star formation, or DWD formation efficiency $\eta_{\rm{form}}(Z)$, varies with metallicity. Consequently, a metallicity-dependent binary fraction further impacts the efficiency of DWD formation within the Galaxy. Figure \ref{fig:form_eff} shows the DWD formation efficiency as a function of metallicity for each DWD type, binary fraction model, and binary evolution parameter variation. In general, the formation efficiency decreases with increasing metallicity. This effect is exaggerated for model FZ which assumes a close binary fraction which also decreases with increasing metallicity. In this section, we give a brief overview of the general trends observed in formation efficiency and their underlying causes. For a detailed description for each DWD type and binary evolution parameter variation, see Appendix \ref{appendix:form_eff}.

There are two common ways that the formation of DWDs from stellar progenitors is inhibited. The first is through stellar mergers, where the binary merges before both components become WDs. The second is due to evolutionary outcomes like forming a different DWD type, or not forming a DWD by the present day. For all DWD types except ONe + X, which has other dominating effects hindering formation efficiency like metallicity-dependent stellar winds, we find that different regimes of ZAMS orbital period leads to different specific evolutionary channels.

In our \cosmic\ models, lower-metallicity stars evolve faster than high-metallicity stars and have larger radii near the end of the MS. This changes the timescales of RLO: low-metallicity stars tend to fill their Roche lobes while still on the MS, meaning mass transfer can remain stable which serves to widen the binary. In contrast, high-metallicity systems initiate mass transfer after the donor has left the main sequence, so the binary is more likely to enter a CE phase. Moreover, high-metallicity progenitors tend to have deeper convective envelopes such that a larger fraction of their mass is contained in the envelope relative to lower-metallicity counterparts of the same mass \citep{Amard2019, Amard2020}. When these systems enter a CE phase, it thus requires more orbital energy to eject the CE than for a low-metallicity star at the same point in its evolution, causing a large amount of orbital shrinking. These effects often lead to a stellar merger at higher metallicities for a system that would have survived the CE phase at a lower metallicity, thus causing a decrease in formation efficiency with metallicity.

The skews introduced by the metallicity-dependent ZAMS orbital period distribution and binary fraction of FZ also contribute to the formation efficiency trends. The log-normal metallicity-dependent $P_{\rm{orb}}$ distribution increasingly skews towards orbital periods below 10$^4$ days for decreasing metallicity. This amplifies the effects of metallicity-dependent stellar evolution discussed above. For example, the skew induces more mergers at high metallicities where the deeper convective envelope leads to increased orbital shrinkage. However, it is also of note that the log-uniform orbital period distribution of F50 produces a higher number of the shortest-orbital period systems (below $\sim100$ days) than FZ, since the tail of the FZ distribution falls quickly below this threshold. These systems also have a propensity for merging before the formation of a DWD. 

Our binary evolution parameter variations emphasizes the interplay of the aforementioned effects of the distributions of F50 and FZ with the physics of binary evolution. For example, the lower CE efficiency of variation $\alpha25$ is the dominating formation efficiency inhibitor since it induces a higher merger rate. In contrast, variations $\alpha5$ and $q3$ relax hindrances to DWD formation (through an increased ability to survive a CE phase and undergo stable mass transfer respectively). We thus find less mergers overall in both cases, and that the effect of the metallicity-dependent $P_{\rm{orb}}$ distribution dominates the formation efficiency trends rather than restrictions imposed by binary evolution prescriptions.

\begin{figure*}
    \centering
	\includegraphics[width=0.93\textwidth]{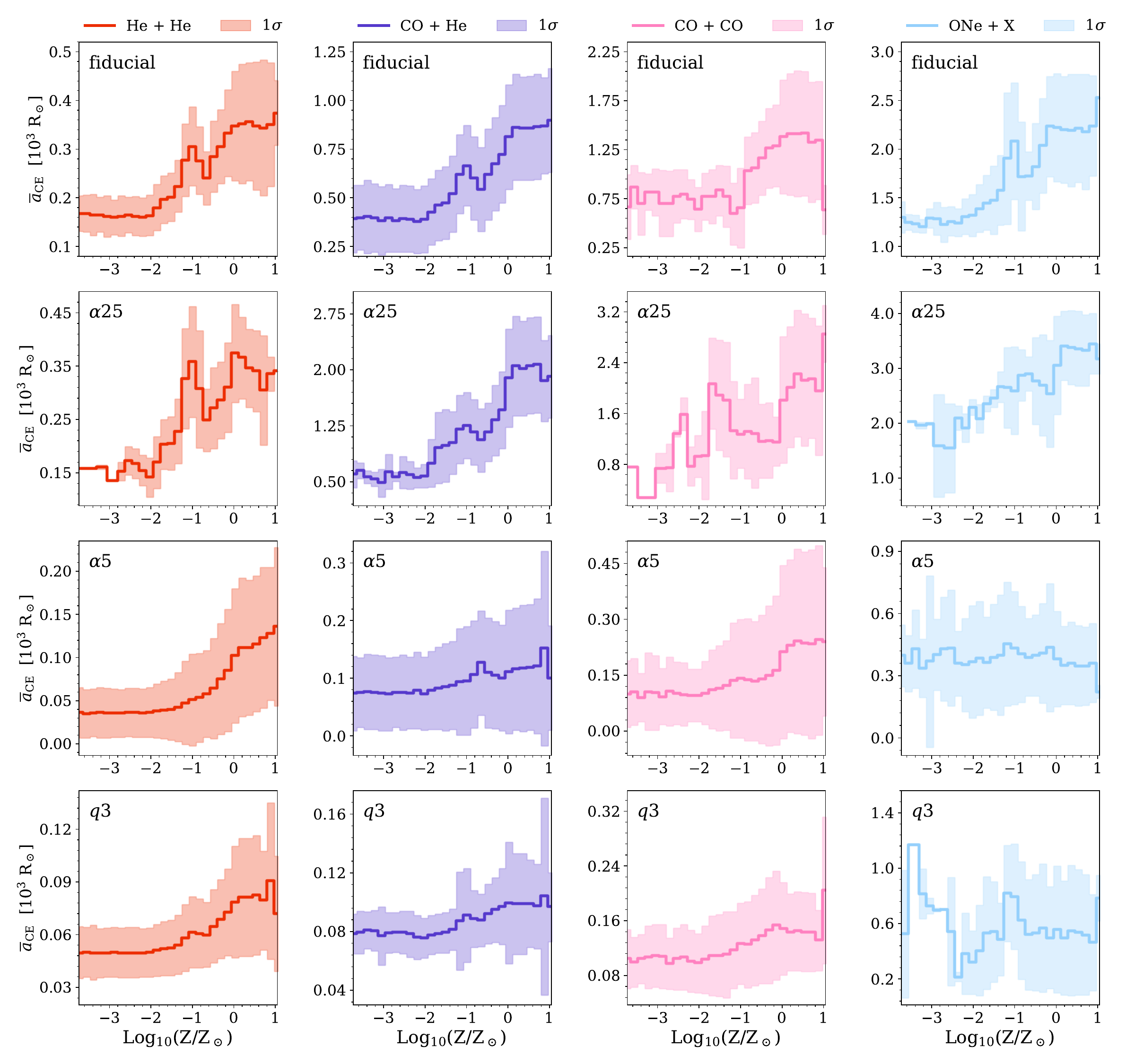}
    \caption{Average interaction separation, $\overline{a}_{\rm{CE}}$ of progenitors of close DWDs across metallicity for each DWD type and variation from model F50. Solid lines show the average separation at the first CE for binaries in each metallicity bin. The shaded regions show the $1\sigma$ spread around the mean within each metallicity bin. The average interaction separation increases with metallicity for every DWD type, except for $\alpha5$ which remains approximately constant for CO + He and ONe + X. The positive trend in the average interaction separation is a direct consequence of larger envelope masses of higher-metallicity donors which are less evolved than their lower-metallicity counterparts. The sharp drop of the CO + CO average at the highest-metallicity bin of the fiducial variation is due to binning effects of the figure: it only contains 304 systems, of which the majority stem from a single low-separation system which was sampled 258 times times during the simulation. The regions of metallicity where there is not shading beyond the mean trend are also due to low numbers of DWDs in each bin.} 
    \label{fig:CEsep}
    \script{CEsep.py}
\end{figure*}

\subsection{Metallicity trends in DWD progenitor common envelope separation}\label{sec:CEsep}

\begin{figure*}
	\includegraphics[width=\textwidth]{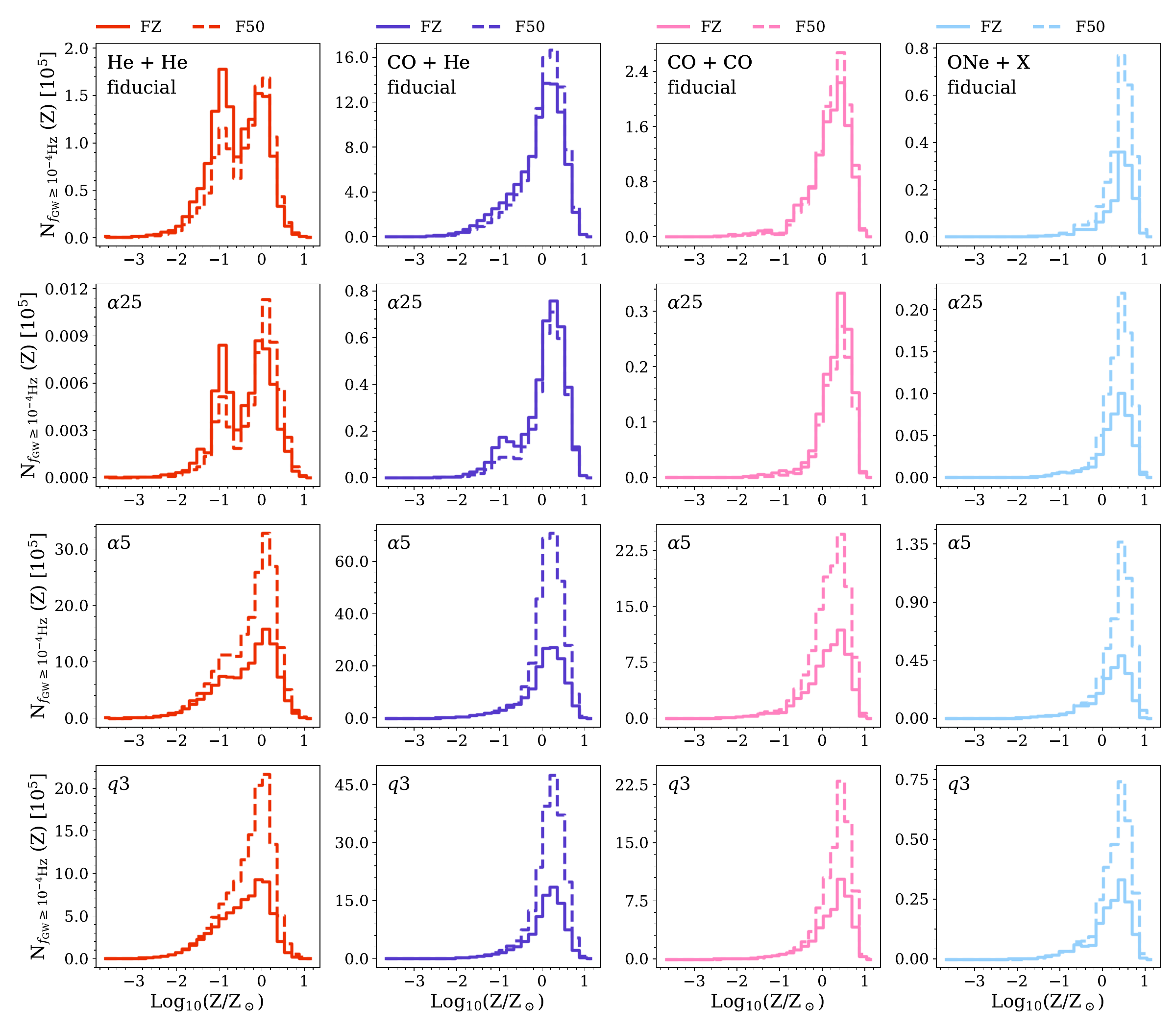}
    \caption{The number of LISA-band systems formed for each DWD type and parameter variation as a function of the base-10 logarithm of metallicity, normalized to solar value. The solid line shows the FZ population with a metallicity-dependent close binary fraction incorporated, and the dashed line shows the F50 population for a standard binary fraction of 0.5. The LISA population of DWDs is dominated by stars with super-solar metallicities. This is true even for model FZ, where $f_{\rm{b}}(Z)$ drops off significantly for higher metallicities, because of the large number of stars formed in \textbf{m12i} beyond $Z\simeq Z_\odot$. There is a double peak in the fiducial and $\alpha25$ He + He populations; the first peak is caused by the sharp drop in formation efficiency past $Z\simeq0.1\,Z_\odot$ which is then greatly overcompensated for by the amount of star formation at higher metallicities which forms the second peak.}
    \label{fig:lisa_nums}
    \script{lisa_nums.py}
\end{figure*}

All systems that end up radiating GWs in the LISA band have undergone at least one phase of CE evolution. For systems which experience a stable RLO mass transfer as the first interaction, the CE phase plays a key role in shrinking systems with initially wide separations to bring them into the LISA band. 

Figure \ref{fig:CEsep} shows the average separation at the first instance of CE evolution, $\overline{a}_{\rm{CE}}$, of all DWD progenitors that result in systems orbiting in the LISA frequency band at present day, as a function of metallicity for each DWD type. The solid lines denote the average value, and the $1\sigma$ variance is shown in the surrounding shading. We show this for our fiducial variation and each of our binary parameter variations. For our fiducial, $\alpha25$ and $\alpha5$ parameter variations, the average CE separation increases in general across metallicity. Higher-metallicity binaries will interact earlier in the binary's lifetime than a lower-metallicity binary of equal separation due to their relatively larger maximum radii. Since higher-metallicity binaries are also more likely to merge during CE interactions because of their relatively more massive donor envelopes, the DWDs which survive and eventually orbit in the LISA band originate from systems with higher interaction separations which allow their orbits to shrink significantly during CE phases without merging. This regulation plays a key role in smearing out any observable effects of a metallicity-dependent binary fraction in the population of DWDs observable by LISA. 

The slope of the increase for these three models is dictated by the CE efficiency $\alpha$. As discussed in Section \ref{sec:formeff}, with increasing metallicity there is a deepening of the convective envelope of the star compared to its lower metallicity counterpart, requiring more orbital energy to be available in order to eject the CE. Thus the high-metallicity binaries must have even wider separations for lower $\alpha$ values to ensure the binary does not merge before the envelope is ejected. We therefore see a steeper separation slope for $\alpha25$, a less steep slope for $\alpha5$, and an intermediate slope for our fiducial variation which has $\alpha=1$.

The DWDs in model $q3$ that survive into the LISA band at present day have different formation channels than those of our other three parameter variations. Most importantly, in contrast to two CE phases for the fiducial, $\alpha25$ and $\alpha5$ parameter variations, the DWD progenitors in parameter variation $q3$ typically undergo a phase of stable mass transfer as the binary's first interaction. During mass transfer the orbit first shrinks, and then will slightly widen once the binary's mass ratio flips. This process leads to formation of the first WD component. The companion is then formed later in the binary's evolution through a CE phase. We show the CE phase which results from the second interaction for parameter variation $q3$ in Figure \ref{fig:CEsep}.

Of particular interest in variation $q3$ is the CO + He systems. For all of the $q3$ CO + He binaries that reside in the LISA band at present, the He WD unexpectedly forms first through stable mass transfer, with the CO WD forming second through a CE. These binaries tend to all have similar resulting orbital separations from mass transfer, as can be seen by the relatively constant curve for CO + He separation at CE onset. There is still a slight increasing trend in CO + He (as well as CO + CO and a more pronounced trend in He + He) for $q3$ that stems from the metallicity-dependent convective envelope discussion above. We also see that the average interaction separations are overall lower for variations $q3$ and $\alpha5$ than for the other two due to the orbital shrinkage that has already occurred from stable mass transfer and the efficient CE phase that will lead to minimal orbital shrinkage respectively.

\section{Metallicity Dependence of the LISA DWD Population}
\label{sec:LISA_met}

While metallicity impacts the intrinsic properties of our simulated DWD populations as described in Sections~\ref{sec:formeff} and \ref{sec:CEsep}, when we consider the present-day Galactic close DWDs we find that the population detectable by LISA largely only changes in total number, but not in frequency distribution. The number of DWDs in the LISA frequency band decreases by $\sim50\%$ when comparing model F50 to model FZ for variations $\alpha5$ and $q3$, but remains approximately unchanged for our fiducial and $\alpha25$ variations.

\begin{figure*}
	\includegraphics[width=\textwidth]{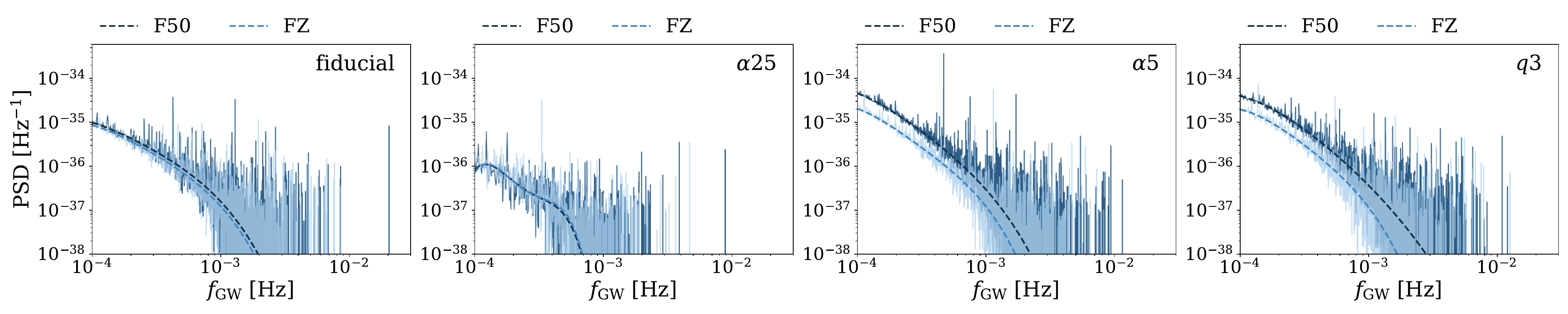}
    \caption{The PSD of the entire Galactic DWD GW foreground, summed over all metallicities and DWD types vs GW frequency for models FZ and F50, where the PSD is downsampled by a factor of $20$ for plotting purposes. Each panel shows a different parameter variation. The vertical lines show the PSD where model F50 is shown in dark blue and model FZ is shown in light blue. The dashed lines show fits to the rolling boxcar median of width 1000 bins for each PSD (see Section \ref{sec:LISA_obs} for a discussion). A metallicity-dependent binary fraction (model FZ) yields fewer DWDs across all frequencies than a $50\%$ binary fraction (model F50) by roughly a factor of two for the $\alpha5$ and $q3$ parameter variations. This produces a lower GW confusion foreground for frequencies with $f_{\rm{GW}}\lesssim 10^{-3}$ Hz. In contrast, the fiducial and $\alpha25$ parameter variations show very similar confusion foregrounds.}
    \label{fig:PSD}
    \script{PSD.py}
\end{figure*}

\begin{figure*}
	\includegraphics[width=\textwidth]{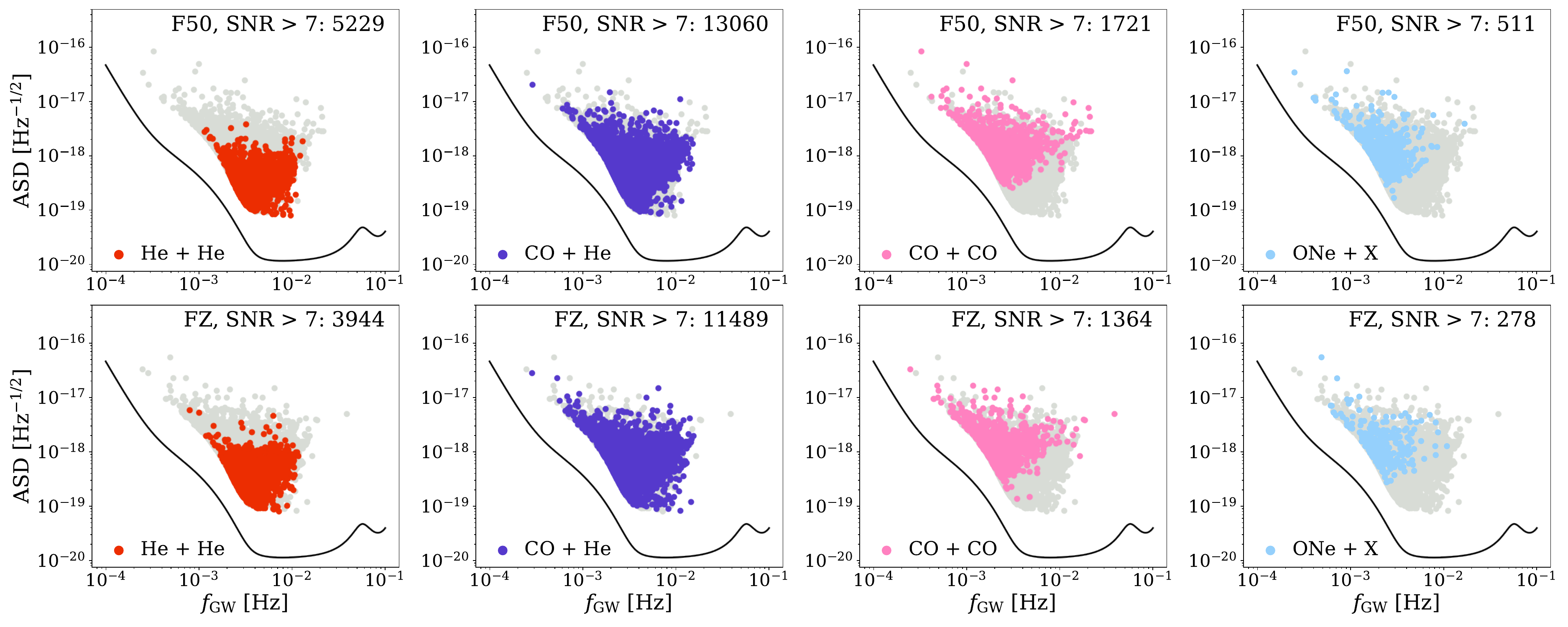}
    \caption{The ASD vs GW frequency for DWDs resolved with SNR $> 7$ for each DWD type where the top row shows the population from model F50 and the bottom row shows the population from model FZ for the fiducial variation. In each panel, the LISA sensitivity curve, including the confusion foreground for each model, is shown in black and the total population for each model is shown in grey. We find that each model qualitatively exhibits similar characteristics and that the only change is in the yield of resolved DWDs for each type based on the strength of the confusion foreground.}
    \label{fig:LISA_SNR}
    \script{LISA_SNR.py}
\end{figure*}

Figure \ref{fig:lisa_nums} shows the number of DWDs orbiting with frequencies $f_{\rm{GW}} > 0.1\,\rm{mHz}$ against metallicity for each DWD type and parameter variation. The solid lines show DWDs from model FZ, and the dashed lines show DWDs from model F50. The He + CO, CO + CO, and ONe + X populations each have strong peaks in the number of DWDs near solar metallicity at which the majority of star formation in galaxy {\bf{m12i}} occurs. The largest contribution to the population comes from metallicities above $\sim 0.01\,Z_\odot$. Any discrepancy between the two binary fraction models is also the most significant above this threshold. When creating our DWD populations the DWD formation efficiency, number of \textbf{m12i} star particles, and the metallicity-dependent binary fraction and ZAMS orbital period distribution all compete. The amount of stars formed in \textbf{m12i} at higher metallicity values overwhelms the drop in DWD formation efficiency by multiple orders of magnitude, so this effect dominates when determining the number of stars initially sampled in the population for $f_{\rm{b}}$. 

There are two peaks in the distribution of He + He DWDs which are most prominent in the fiducial and $\alpha25$ variations, and not as obvious for $\alpha5$ or $q3$. The prominent first peak in fiducial and $\alpha25$ occurs because near $Z\simeq 0.1\,Z_\odot$, the DWD formation efficiency transitions from near constant values to a sharp decrease (see Figure \ref{fig:form_eff}). However, for a drop in the formation efficiency by a factor of of $\sim$ six, the amount of star formation in galaxy \textbf{m12i} increases by more than an order of magnitude for supersolar metallicities. Above $Z=Z_\odot$ this overcompensates for the efficiency drop, producing the second peak. For $\alpha5$ and $q3$, the lack of dramatic orbital shrinking -- from the increased CE efficiency in the former and increased propensity for stable mass transfer in the latter -- allow for less failed DWDs due to stellar mergers. This leads to less dramatic drops in formation efficiency across metallicity which thus suppresses the formation of two peaks in the number of LISA-band systems.

The $\alpha5$ and $q3$ variations produce a higher number of F50 systems than FZ when summed over all DWD types. This is due to the log-uniform $P_{\rm{orb}}$ distribution of model F50 having more of the shortest-period binaries than the tails of the log-normal distribution of FZ; a discrepancy which becomes more significant with increasing metallicity as the FZ distribution moves from a peak of log$_{10}$($P_{\rm{orb}}\rm{/day})\simeq 3-4$ to log$_{10}$($P_{\rm{orb}}\rm{/day})\simeq 4-6$. This tail is less likely to survive to the LISA band at present for the fiducial and $\alpha25$ variations due to the increased propensity for mergers of DWD progenitors. This effect produces an equal number of systems between FZ and F50 for the fiducial and $\alpha25$ variations. Although the close binary fraction leads to a greater number of low-metallicity systems, this is compensated for by the FIRE star particle metallicity distribution as well as the tendency for the shortest-period DWD progenitors in model F50 to merge, essentially washing out all differences between the populations.

\begin{figure*}
	\includegraphics[width=\textwidth]{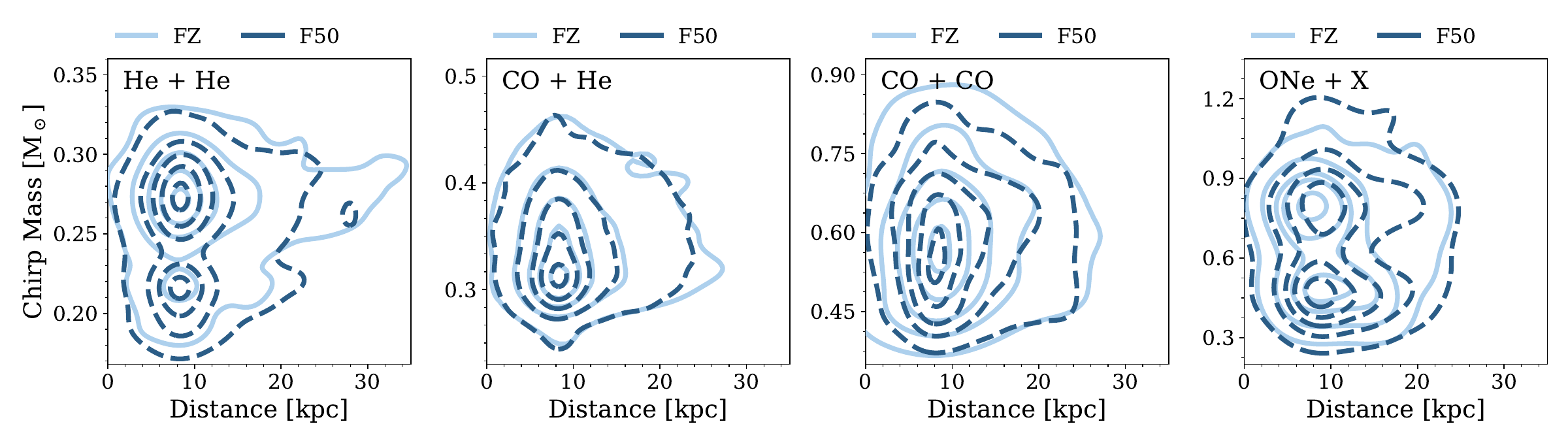}
    \caption{The chirp mass vs distance for each DWD type is shown for our fiducial parameter variation. Only systems with observable evolution in their GW frequency, i.e those which are chirping, and with SNR $>7$, are shown, since these are systems for which distance can be separated from chirp mass within their strain amplitude. Each panel shows one DWD type, summed over all metallicities. Model FZ is indicated with solid light blue contours, and model F50 is indicated with dark blue dashed contours respectively. Contours are shown at the $5^{\rm{th}}$, 25$^{\rm{th}}$, 50$^{\rm{th}}$, 75$^{\rm{th}}$, and 95$^{\rm{th}}$ percentiles. Despite intrinsic changes to population properties induced by a metallicity-dependent binary fraction and a reduction in the height of the DWD Galactic foreground, the distributions are very similar.}
    \label{fig:Mc_vs_dist}
    \script{Mc_vs_dist.py}
\end{figure*}

A change in the number of DWDs in model FZ relative to model F50 is also apparent in the GW PSD LISA will observe. We show the GW PSD of each model, as well as the confusion forground estimate, in Figure~\ref{fig:PSD}. We downsample the PSD frequencies by a factor of 20 for plotting purposes. While model F50 (dark blue) and model FZ (light blue) produce several thousand large spikes in the PSD across LISA's frequency band, the overall foreground height, including the confusion, is larger for model F50 for parameter variations $\alpha5$ and $q3$. This is a direct consequence of the overall reduction in the size of the close DWD population in model FZ. The foreground height remains unchanged for the fiducial and $\alpha25$ variations when comparing models FZ and F50.

Similar to previous studies, we find that LISA will be able to resolve several thousand DWDs. Figure \ref{fig:LISA_SNR} shows the amplitude spectral density vs GW frequency of the resolved systems with SNR $> 7$ for each DWD type, where the top and bottom rows show results for models FZ and F50 respectively for the fiducial parameter variation. For comparison, we also show the LISA sensitivity curve, including the modeled confusion foreground from each population in black, and the entire population for each model in grey. Plots for our three parameter variations and discussion on the impact of each varation on the resolved populations can be found in Appendix \ref{appendix:ASD_vars}. Apart from each DWD type having a different abundance of resolved systems, the population-wide characteristics remain unchanged between the two binary fraction models. The populations containing at least one He WD occupy the lower-ASD, higher-GW frequency region of parameter space compared to the total population, with CO + He DWDs having larger ASDs than the He + He DWD population. Conversely, DWD types without a He WD component tend to occupy the higher-ASD, lower-GW frequency region of parameter space. This difference is largely due to the formation scenario of DWDs containing a He WD, which form from the ejection of a common envelope created by the He WD progenitor. These lower-mass progenitors overflow their Roche lobes at closer separations relative to the higher-mass progenitors (e.g. Figure~\ref{fig:CEsep}) and thus also produce closer DWDs. While the distance to any one DWD strongly influences its ASD, DWD populations without a He WD component have, on average, higher ASDs due to their more massive WD components.

The distance and chirp mass of DWDs which exhibit observable orbital evolution due to the emission of GWs during the LISA mission can be measured. This is because the chirp mass -- distance degeneracy in the observed strain can be broken with the observed GW frequency evolution, or chirp. Assuming a chirp resolution of $1/T_{\rm{obs}}^2 \sim 8\times10^{-9}\,\rm{Hz}^2$, we select the DWDs whose chirp masses and distances can be measured. Figure \ref{fig:Mc_vs_dist} shows the chirp mass vs the luminosity distance for each DWD type in this selected population for the fiducial variation. The same figure for our other three parameter variations and accompanying discussion is shown in Appendix \ref{appendix:Mc_dist_vars}. The contours show the $5^{\rm{th}}$, $25^{\rm{th}}$, $50^{\rm{th}}$, $75^{\rm{th}}$, and $95^{\rm{th}}$ percentiles for models FZ (light blue) and F50 (dark blue, dashed). Despite the reduction in the height of the confusion foreground when considering model FZ relative to F50, we find that LISA is unable to differentiate between the chirp mass -- distance distributions of the two models. 

\section{Galactic DWD Confusion Foreground Discussion}
\label{sec:model_compare}

\begin{figure}
	\includegraphics[width=0.45\textwidth]{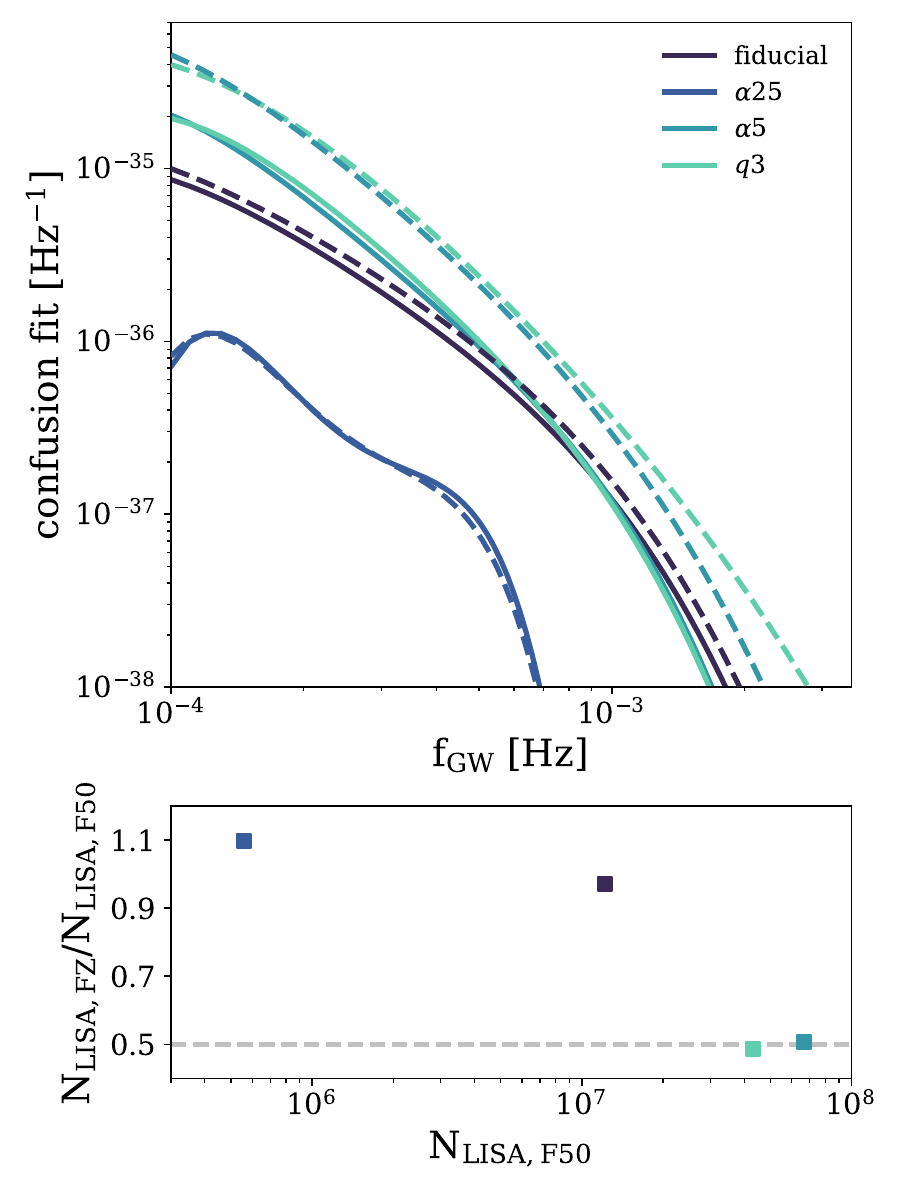}
    \caption{The top panel shows the Galactic DWD confusion fit vs GW frequency for different binary evolution parameter variations (colors) with model F50 shown in dashed lines and model FZ shown in solid lines. The bottom panel shows the ratio of the number of DWDs orbiting in the LISA frequency band for model FZ to model F50 vs to number of DWDs orbiting in the LISA band for model F50 only. While both the height of the confusion foreground and number of LISA DWDs changes for each variation, the FZ models within each variation exhibit either very little change or a  near-constant reduction by a factor of two compared with the F50 models.}
    \label{fig:model_comp}
    \script{model_comp.py}
\end{figure}

In each parameter variation, changing the binary evolution assumptions dramatically changes the formation and evolution of the DWD populations. These changes lead to large shifts in the overall number of close DWDs in our synthetic present-day Milky Way-like galaxies. We find that the total number of DWDs with $f_{\rm{GW}}>10^{-4}\,\rm{Hz}$ increases for both variations $q3$ and $\alpha5$ with respect to the fiducial case. This is because there are fewer stellar mergers which occur before the formation of a DWD, thus allowing more systems to evolve due to GW emission and orbit in the LISA frequency band at present. Conversely, for variation $\alpha25$, we find that the number of DWDs orbiting with frequencies in the LISA band is drastically reduced. This is because of the highly inefficient use of orbital energy to eject the common envelope, which produces more stellar mergers, or closer binaries which are more prone to future mergers.

Interestingly, when we compare the populations of each binary fraction model for our variations, we find that the number of close DWDs remains approximately unchanged for our fiducial set of assumptions and $\alpha25$ variation, but is reduced by 50\% for variations $\alpha5$ and $q3$. This is illustrated in Figure \ref{fig:model_comp}. The top panel shows the confusion foreground fits for each binary evolution variation (different colored lines) and for each binary fraction model where the solid lines show FZ models and dashed lines show F50 models. The bottom panel shows the ratio of the number of DWDs orbiting in the LISA frequency band for the FZ models vs the F50 models for each variation. The number of DWDs in the LISA band spans over two orders of magnitude, but the ratio of FZ to F50, as well as the spectral shape of the confusion fit, is either relatively unchanged or reduced by a factor of approximately two. This suggests that assuming a metallicity-dependent binary fraction may change the size of the Galactic close DWD population by a factor of $\sim$ two and the strength of the Galactic DWD GW foreground for LISA depending on the chosen binary evolution model. We highlight, however, that for variations where the height of the confusion foreground is reduced, the number of resolved sources is not reduced to an equal degree. This is because fewer competing GW signals from DWDs in the LISA frequency band leads to more individually resolved DWDs. See Appendix \ref{appendix:ASD_vars} for more details.

\section{Conclusions}\label{sec:conclusions}
In this study, we have investigated the effects of assuming a metallicity-dependent binary fraction on the formation and evolution of the Galactic population of DWDs with a special focus on the implications for LISA. Based on our synthetic Milky Way-like galaxy catalogs of DWDs, we find that applying a metallicity-dependent binary fraction changes the formation efficiency and evolutionary history of DWD populations. However, when considering the close DWD populations observable by LISA, we find that the only distinguishing features between models which assume a metallicity-dependent binary fraction (model FZ) and models which assume a flat $50\%$ binary fraction (model F50) are the population sizes and the strength of the Galactic DWD GW foreground. Models which assume a metallicity-dependent binary fraction produce Galactic DWD populations that were unchanged relative to the standard model assumptions. However, our results are highly dependent on the chosen binary evolution model. We extended our study to include three binary evolution parameter variations to investigate the sensitivity of the DWD population changes in assumptions for mass transfer stability and common envelope ejection efficiencies. While our binary evolution parameter variations change the size of the LISA-observable populations dramatically, the size of the close Galactic DWD population and the height of the confusion foreground for models which assume a metallicity-dependent binary fraction appears to either remain unchanged \emph{or} be reduced by 50\%. These two results highlight the interplay between the dominating factors which shape the LISA DWD population: the metallicity-dependent star formation history of the \textbf{m12i} galaxy, the metallicity-dependent close binary fraction which alters the ZAMS orbital period distribution, and the impact of binary evolution assumptions on DWD formation. This comparison emphasizes the sensitivity of the Galactic population of LISA-detectable DWDs to the intricacies of binary evolution. Binary evolution assumptions like CE efficiency are undergoing continuous study in current literature and an exact value has yet to be identified. By exploring a range of possibilities in this work, we capture the variation in predictions of the LISA DWD population that may be observed.

An important consequence of a lower Galactic DWD confusion foreground is that relative to the total DWD population, more DWDs can be individually resolved because of the reduction in competing GW signals. While the number of DWDs radiating GWs in the LISA frequency band is reduced by a factor of two for model FZ relative to model F50 for variations $\alpha25$ and $q3$, the number of resolved sources is less affected with a population-wide reduction of $22\%$ for the former, and surprisingly only by 5\% for the latter. These results are far-reaching since the strength of the Galactic DWD confusion foreground has direct consequences on the detectability of all other LISA sources with small SNRs. An increase in resolution capability from the reduced confusion foreground can be extended to other Galactic binaries that LISA will observe at these frequencies like those involving neutron stars and stellar-origin black holes, as well as other more exotic GW sources like merging supermassive black holes, extreme mass ratio inspirals, or cosmological GW backgrounds. Based on our results, we suggest that studies which employ fits to the confusion foreground based on population synthesis results consider reducing the strength of the Galactic foreground PSD by a factor of two depending on the chosen binary evolution model.

\begin{acknowledgments}
The authors are grateful for helpful discussions with Carles Badenes, Christine Mazzola Daher, Kaitlin Kratter, and the Gravitational Waves and Astronomical Data groups at the CCA. The authors are also grateful to the referee for providing a thoughtful review which strengthened the presentation of the manuscript. S.T.\ was supported by an Undergraduate Student Research Award (USRA) at CITA from the Natural Sciences and Engineering Research Council of Canada (NSERC), Reference \# 498223. K.B.\ is grateful for support from the Jeffrey L. Bishop Fellowship. The Flatiron Institute is supported by the Simons Foundation. This research was supported in part by the National Science Foundation under Grant No. NSF PHY-1748958.
\end{acknowledgments}

\section*{Data Availability}

All data and software required to reproduce our results are available through GitHub and Zenodo which are accessible through the icon links associated with our abstract and each figure.

\software{\texttt{astropy} \citep{astropy:2013, astropy:2018}; 
          \cosmic\ \citep{Breivik2020a};
          \legwork\ \citep{Wagg2021};
          \texttt{matplotlib}\ \citep{matplotlib}; 
          \texttt{numpy}\ \citep{numpy}; 
          \texttt{pandas}\ \citep{mckinney-proc-scipy-2010, reback2020pandas}; 
          \texttt{scipy}\ \citep{scipy}
          \texttt{seaborn}\ \citep{Waskom2021}
          }
          
\bibliographystyle{aasjournal}
\bibliography{bib}{}

\appendix
\section{Formation efficiency trends}\label{appendix:form_eff}
As discussed in Section \ref{sec:formeff}, the formation efficiency of DWDs exhibits a metallicity dependence. This is due to the impact of metallicity on both stellar evolution, and the ZAMS orbital period distribution of model F50 vs. FZ. Each DWD type and binary evolution parameter variation exhibits unique trends in efficiency and in their evolutionary channel. In the following subsections we go through each DWD type in detail, first for our fiducial simulations and then for our variations.

\subsection{Fiducial formation efficiency trends}\label{subsec:formeff_fid}

\subsubsection{He + He}\label{formeff_HeHe}
For He + He DWDs, the sudden drop in formation efficiency near $\log(Z/Z_{\odot})=-1.0$ (see Figure \ref{fig:form_eff}) is generally caused by the timescale for which the initially more massive star in the DWD progenitor overflows it's Roche lobe. As discussed in Section \ref{sec:formeff}, at lower metallicities, donors tend to fill their Roche lobes while they are still on the MS and the mass transfer remains stable and serves to widen the binary. At higher metallicities, mass transfer is initiated when the donor has left the main sequence, and the binary enters a common envelope evolution.

The ZAMS orbital period leads to different specific He + He evolutionary channels. For short-period systems with periods below $10~\rm{days}$, a CE phase leads to a stellar merger due to insufficient orbital energy to eject the envelope. Stellar mergers continue to dominate the evolutionary pathways of systems with intermediate orbital periods ($1 < \log_{10}(P_{\rm{orb}}/\rm{day}) < 2.5$). However, an additional growing number of merging systems arises with one WD component and one stellar companion, and surviving systems which don't interact at all and thus do not form a He + He DWD. The distinction between the various scenarios in this intermediate orbital period range depends on the combination of their ZAMS masses and orbital periods. Finally, at wider initial periods $\log_{10}(P_{\rm{orb}}/\rm{day}) > 2.5$, the decrease in formation efficiency is dominated by systems which never interact and thus do not form a DWD before the present day.

\subsubsection{CO + He}\label{formeff_COHe}
In general, a CO + He DWD fails to form because either a stellar merger occurs before DWD formation, or a CO + CO or He + He DWD is formed instead. We discuss the specifics of these effects for three ZAMS orbital period regimes below. Model FZ yields higher formation efficiencies than F50 because CO + He DWDs prefer to form from progenitor binaries with orbital periods near $\log_{10}(P_{\rm{orb}}/~\rm{day})\sim2-4$, which is where the orbital period distribution of FZ peaks.

For short-period CO + He DWD progenitor binaries with orbital periods below $\sim30~\rm{days}$, the only channel for stellar mergers before DWD formation is during a CE evolution. This channel is similar to the stellar merger channel for He + He DWDs, but due to higher-mass progenitors the CE evolution results in the binary component with a higher mass becoming a CO WD. At higher metallicities, the CO WD merges with its companion. As discussed in Section \ref{sec:formeff}, higher-metallicity progenitors have a larger fraction of their mass in the convective envelope when compared to lower-metallicity stars of the same mass. Thus a CE phase with a higher-metallicity progenitor requires more orbital energy to eject the common envelope, causing a larger amount of orbital shrinking which results in a merger later in its binary evolution.

For CO + He DWD progenitor systems with intermediate orbital periods ($1.5\leq \log_{10}(P_{\rm{orb}}/\rm{day})< 2.5$) the mechanisms which impact formation efficiency are complex. The dominant way CO + He DWDs fail to form is stellar mergers that occur during the second CE phase. Low-metallicity systems survive this second CE phase whereas high-metallicity systems do not. Similar to the short-period systems, there is increased orbital shrinkage for high-metallicity progenitor binaries due to the deeper convective envelope of the CE evolution donor. This creates shorter post-CE orbital periods which leads to mergers during the second CE phase.

If a lower-metallicity system forms a CO + He DWD and a higher-metallicity system does not, this could also be because the initially more massive binary component initiates a CE while on the giant branch instead of the asymptotic giant branch. This leaves behind a He WD with a stellar companion instead of a CO + He, which is the dominating scenario that restricts formation efficiency. There are a few edge cases where either a He + He DWD is formed instead, or a CO WD and a stellar companion is formed and there has been very nearly but not quite enough time for a CO + He DWD to form. 

Long-period binaries with $\log_{10}(P_{\rm{orb}}/\rm{day})>2.5$ also display complex scenarios that hinder CO + He DWD formation. In near-equal contributions, our \cosmic\ simulations produce either stellar mergers or stable non-DWD binaries at the end of the Hubble time. At these orbital periods, stellar mergers always occur with a CE phase between a stellar companion and a CO or He WD. Again, the mergers occur because of increased CE donor envelope masses at higher metallicities. A subdominant channel of stable He + He DWDs can also occur when a high-metallicity primary overflows it's Roche lobe while still on the giant branch and thus forms a He WD.

\subsubsection{CO + CO}\label{formeff_COCO}
The decrease in the CO + CO DWD formation efficiency with increasing metallicity stems from different evolutionary channels which arise at the ZAMS orbital period boundary of $\log_{10}(P_{\rm{orb}}/\rm{day})\sim3$. We find that for binaries with orbital periods below this boundary, the most common way that CO + CO DWDs form at lower metallicities but not at higher metallicities is through stellar mergers during a CE phase with a donor that is still on the giant branch. For lower-metallicity binaries, which evolve on faster timescales, the primaries enter CE evolution while on the asymptotic giant branch instead and the binary is able to survive. For binaries with orbital periods above the boundary, the vast majority of systems with wide initial orbits end up as stable binaries. The wide systems which don’t form a CO DWD at high metallicity do so because one or both of the binary components initiate a CE phase while still on the giant branch, thus producing a CO + He or He + He DWD. 

\subsubsection{ONe + X}\label{formeff_ONe}
The strongest effect which hinders formations of higher-metallicity ONe + X DWDs is the strength of metallicity-dependent stellar winds assumed in our model \citep{Vink2001}. The strength of line-driven winds varies more strongly for the more massive ($\geq5\,M_\odot$) progenitors of ONe WDs relative to the other lower-mass WD progenitors. At higher metallicities, ONe WD progenitors can lose enough mass through winds such that they don't ignite their CO cores and thus leave behind a CO WD. Conversely, the lower-metallicity progenitors retain enough mass to cause carbon ignition and leave behind an ONe WD.

\subsection{Parameter variations formation efficiency trends}\label{subsec:formeff_vars}

As discussed in Section \ref{sec:formeff}, our simulations with binary evolution parameter variations show competition between the physics of binary evolution and skews introduced by the metallicity-dependent orbital period distribution and binary fraction. Two important features of our FZ and F50 models drive the differences in formation efficiency for our binary evolution parameter variations. First, the log-normal metallicity-dependent ZAMS orbital period distribution increasingly skews towards $P_{\rm{orb}}\lesssim 10^4$ days for decreasing metallicity. However, second, the log-uniform $P_{\rm{orb}}$ distribution of F50 produces a higher number of the shortest-period systems ($P_{\rm{orb}}\lesssim 100$ days) than for FZ as the tail of the FZ distribution falls below F50 in this regime. Because there are more ZAMS binaries with $P_{\rm{orb}} \lesssim 10^4$ days in F50, as the population evolves there are more likely to be binary interactions in the population for F50 than FZ. This leads to more mergers in $\alpha25$ occurring for F50 before formation of the DWD, and thus a lower $\alpha25$ formation efficiency overall for F50 in contrast to FZ. Below we summarize how the formation efficiency of each DWD type is impacted by the parameter variations. Across all DWD types, we find that there are more mergers for variation $\alpha25$ compared with the fiducial variation, and fewer for $\alpha5$, due to the CE efficiency's impact on CE phase orbital energy requirements. We also find fewer mergers for $q3$, because of the ability for binaries to maintain stable mass transfer, thus increasing their chance of survival.

\subsubsection{Parameter variations: He + He}\label{formeff_HeHe_vars}
For binary systems leading to He + He DWDs, as stated above there are many more mergers for $\alpha25$ than for the fiducial variation. The lower CE efficiency leads to more DWD progenitor mergers from a failed CE ejection, or the CE shrinking the orbit such that it then merges on the second CE phase. As discussed above, there are also more close binaries in the orbital period distribution of F50 than FZ, creating a higher propensity for binary interactions and thus mergers for $\alpha25$. This is illustrated in the $\alpha25$ He + He plot of Figure \ref{fig:form_eff}. Similar to the fiducial parameter variation, a binary that becomes a He + He DWD at lower metallicities would merge before becoming one at higher metallicities because of the longer evolution timescales. These later interactions lead to closer post-CE separations and thus more mergers. The dip at [Fe/H] $\simeq -1.5$ (see Figure \ref{fig:form_eff}) stems from mergers before the DWD forms, again similar to the fiducial parameter variation. The peak that occurs in $\alpha25$ at $\sim 10\%\,Z_\odot$ arises from binaries evolving faster, and thus interacting earlier, as metallicity increases. This means that equivalent binaries that would have merged or formed a CO + He DWD instead at lower metallicities now survive to produce a He + He DWD. At higher metallicities, there are also a higher proportion of wide ZAMS binaries, which evolve off the Main Sequence too slowly to form a He + He DWD, adding to the efficiency drop.

For $\alpha5$, there are significantly fewer DWD progenitor mergers since binaries are more likely to survive a CE phase due to the higher CE ejection efficiency. Systems that would merge for $\alpha25$ instead survive, shrinking to short orbits through CE interactions to reside in the LISA band at present. The number of DWDs then makes up a higher fraction of the total simulated binary population, thus we find in general higher formation efficiencies than for the fiducial parameter variation which sets $\alpha=1$. This effect is even more pronounced in comparison with the $\alpha25$ variation which sets $\alpha=0.25$.

Beyond the effects described above, the efficiency drop across metallicity is similar to the fiducial parameter variation in that it depends on the metallicity-dependent ZAMS orbital period distribution emphasizing short-period systems at low metallicities, and on the change of formation channel for different period ranges across metallicity. We find that variation $q3$ yields the highest formation efficiency. Only the shortest-period binaries do not form a DWD, which are lost through mergers at higher metallicities. At low metallicities, even binaries with ZAMS orbital periods $P_{\rm{orb}}\lesssim 100$ days form He + He DWDs, with binaries in model FZ reaching ZAMS orbital periods less than a day in length. This leads to a higher formation efficiency from F50 over FZ for parameter variation $q3$ because of the higher number of shortest-period ZAMS binaries in the log-uniform distribution of F50.

\subsubsection{Parameter variations: CO + He}\label{formeff_COHe_vars}
For CO + He binaries, there are again many more mergers for parameter variation $\alpha25$ relative to the fiducial case. This is the dominant reason that progenitors do not evolve to produce DWDs at higher metallicities. At low metallicities, a number of progenitor binaries form CO + CO DWDs instead of He + CO DWDs. This is especially the case near $\log_{10}(Z/Z_\odot)\sim -2$, where RLO occurs late enough in a binary's evolution that mass transfer remains stable and CO + CO's are formed rather than CO + He, hence producing a dip in the formation efficiency (see Figure \ref{fig:form_eff}).

Because parameter variation $\alpha5$ yields fewer mergers, the formation efficiency is mostly governed by the metallicity-dependent ZAMS orbital period distribution. At higher metallicities, the majority of cases in model FZ where progenitors do not produce a CO + He DWD are from progenitor mergers, or from He + He DWDs forming instead of CO + He DWDs. At higher metallicities, the ZAMS orbital period distribution of FZ also drops below that of F50 and extends out to orbital periods beyond what would form a CO + He. This causes the formation efficiencies for FZ and F50 to cross at log$_{10}(Z/Z_\odot)\sim-0.5$. This is also apparent for the $q3$ variation, and to a lesser extent in the fiducial case. In $q3$, for any given orbital period we find that lower-mass DWDs originate from lower metallicities, thus leading to an approximately constant formation efficiency for model F50.

\subsubsection{Parameter variations: CO + CO}\label{formeff_CO_vars}
For CO + CO, across all variations, very close ZAMS binaries ($P_{\rm{orb}}\sim\mathcal{O}(100$ days)) at high metallicities form a CO + He or He + He DWD instead of a CO + CO because the binary interacts before core Helium burning of one or both components. There are again many more mergers in $\alpha25$, and fewer mergers for $\alpha5$. Due to the high CE efficiency of $\alpha5$, systems with shorter ZAMS orbital periods survive to become CO + CO DWDs rather than merging or becoming CO + He or He + He DWDs. There are fewer mergers overall for $q3$ than all the $\alpha$ variations because the first RLO occurrence leads to stable mass transfer from the primary to the secondary rather than multiple CE phases. For both the $\alpha5$ and $q3$ variations, the majority of CO + CO DWDs originate in wide enough orbits that they do not interact before DWD formation. This leads to an approximately constant formation efficiency across metallicity in both cases for model F50.

\subsubsection{Parameter variations: ONe + X}\label{formeff_ONe_vars}
For model FZ, the ONe + X formation efficiencies follow similar trends aross each parameter variation. Their formation scenarios are similar to those described in Section \ref{sec:formeff} for ONe + X DWDs in the fiducial binary evolution parameter set. The main difference between each parameter variation is only in the number of sources produced since the number of stellar mergers that occur before DWD formation increases with decreasing common envelope efficiency. In model F50, the formation efficiency is decreased with respect to FZ for parameter variation $\alpha25$ because the log-uniform orbital period distribution produces many ZAMS binaries with very short orbital periods which lead to stellar mergers during the CE phases. The F50 formation efficiency distributions are slightly flattened compared to model FZ for variations $\alpha5$ and $q3$, similar to CO + CO as discussed above.

\section{Amplitude Spectral Densities for Binary Evolution Parameter Variations}\label{appendix:ASD_vars}

\begin{figure*}[h]
    \centering
	\includegraphics[width=0.8\textwidth]{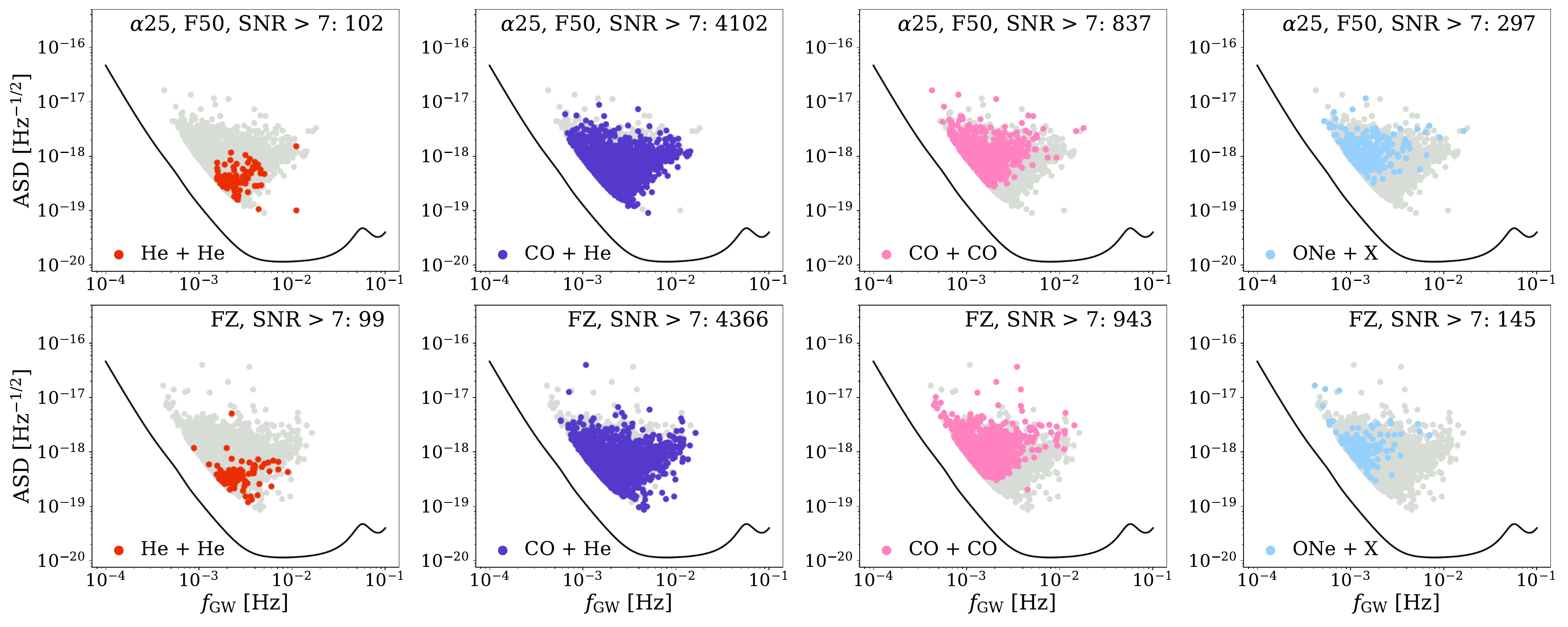}
	\includegraphics[width=0.8\textwidth]{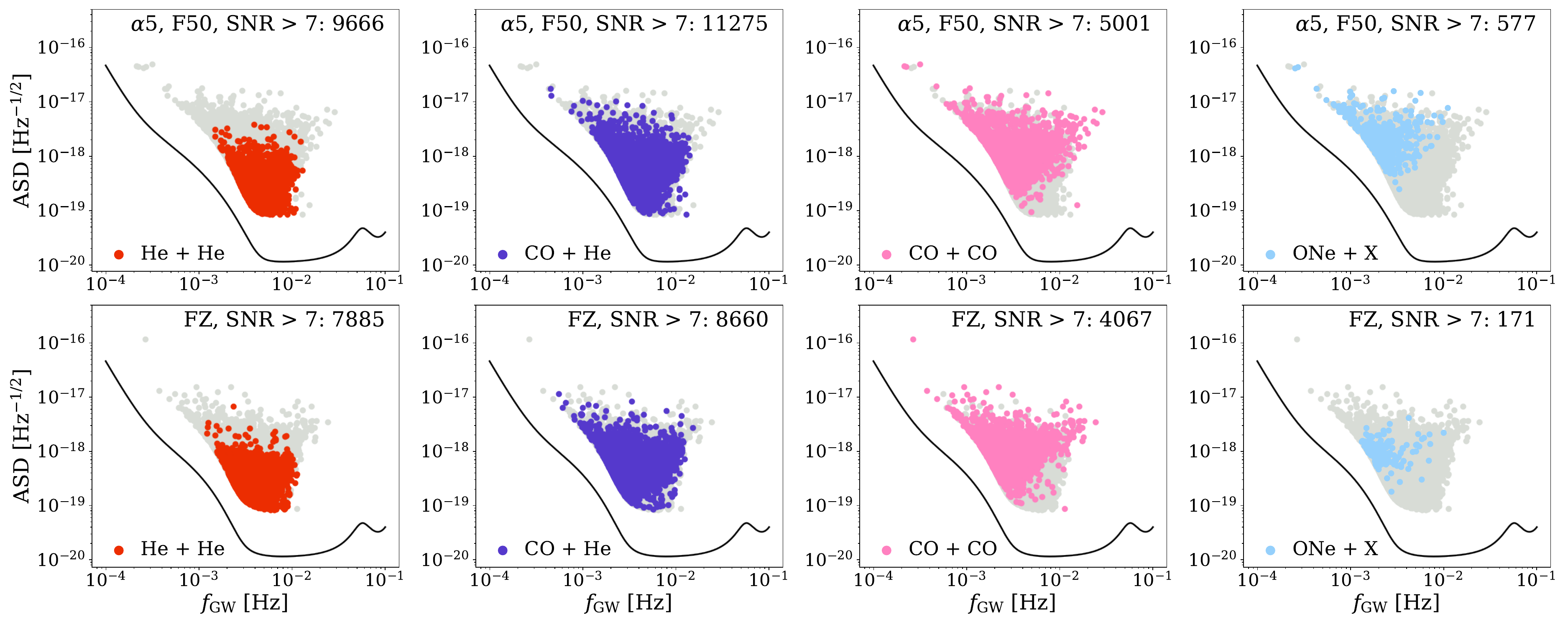}
	\includegraphics[width=0.8\textwidth]{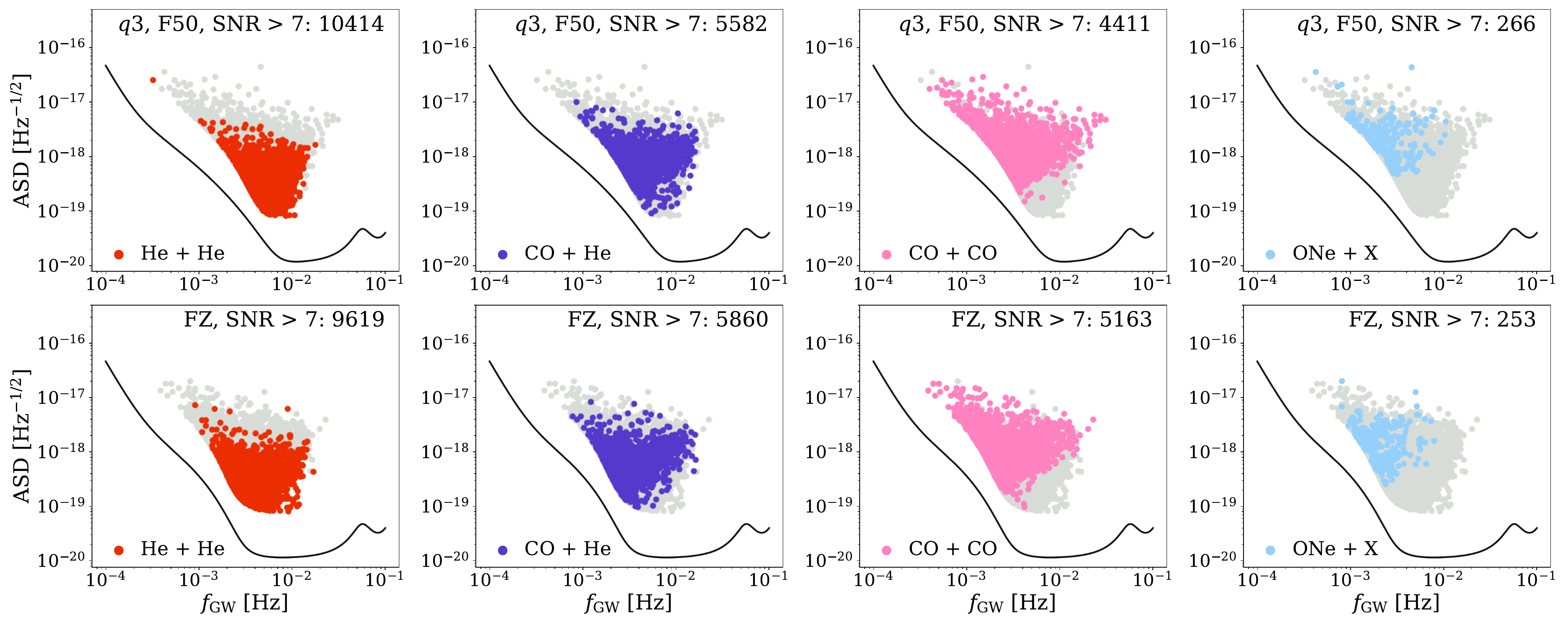}
    \caption{The ASD vs GW frequency for DWDs resolved with SNR $> 7$ for each DWD type and parameter variation. The rows alternate between model F50 and model FZ. In each panel, the LISA sensitivity curve, including the confusion foreground for each model, is shown in black and the total population for each model is shown in grey. Again, we find that each model qualitatively exhibits similar characteristics and that the only change is in the yield of resolved DWDs for each type based on the strength of the confusion foreground.}
    \label{fig:LISA_SNR_vars}
     \script{LISA_SNR.py}

\end{figure*}

Figure \ref{fig:LISA_SNR_vars} shows the ASD vs GW frequency of the resolved systems with SNR $> 7$ for each DWD type, where the rows alternate between models F50 and FZ for the $\alpha25$, $\alpha5$, and $q3$ parameter variations. The LISA sensitivity curve is also shown in black, which includes the modeled confusion foreground (see Section \ref{sec:LISA_met} for details on fitting the foreground) for the given variation. The entire population is shown as grey scatter points. As discussed in Section \ref{sec:LISA_met}, apart from each DWD type having a different abundance of resolved systems, the population-wide characteristics remain unchanged between the two binary fraction models.

The height of the confusion foreground and the number of DWDs radiating GWs with $f_{\rm{GW}} > 10^{-4}\,\rm{Hz}$ affects the number of resolvable systems in each parameter variation. For the fiducial and $\alpha25$ variations, the number of resolvable systems with SNR $>7$ are reduced by $\sim 17$\% and increased by $\sim3\%$ respectively, with relatively unchanged overall LISA population sizes. For $\alpha5$ the number of resolvable DWDs is reduced by only $\sim 22\%$, and the number of resolvable DWDs $q3$ remains almost \emph{unchanged}, with a $\sim 5\%$ reduction even though the overall LISA population of both variations is halved. The majority of systems which become resolvable with a lower foreground (in model FZ) are non-chirping systems. This is because both chirping and non-chirping systems have their populations halved, but the lower foreground allows a relative increase in the number of resolved non-chirping sources. Thus, the incorporation of a metallicity-dependent binary fraction reduces the number of chirping binaries relative to non-chirping binaries by a factor of two for variations $\alpha5$ and $q3$. The number of chirping systems in the fiducial and $\alpha25$ variations remain unchanged between models F50 and FZ.

\section{Chirp Mass -- Distance Distributions for Binary Evolution Parameter Variations}\label{appendix:Mc_dist_vars}

\begin{figure*}[h]
    \centering
	\includegraphics[width=0.8\textwidth]{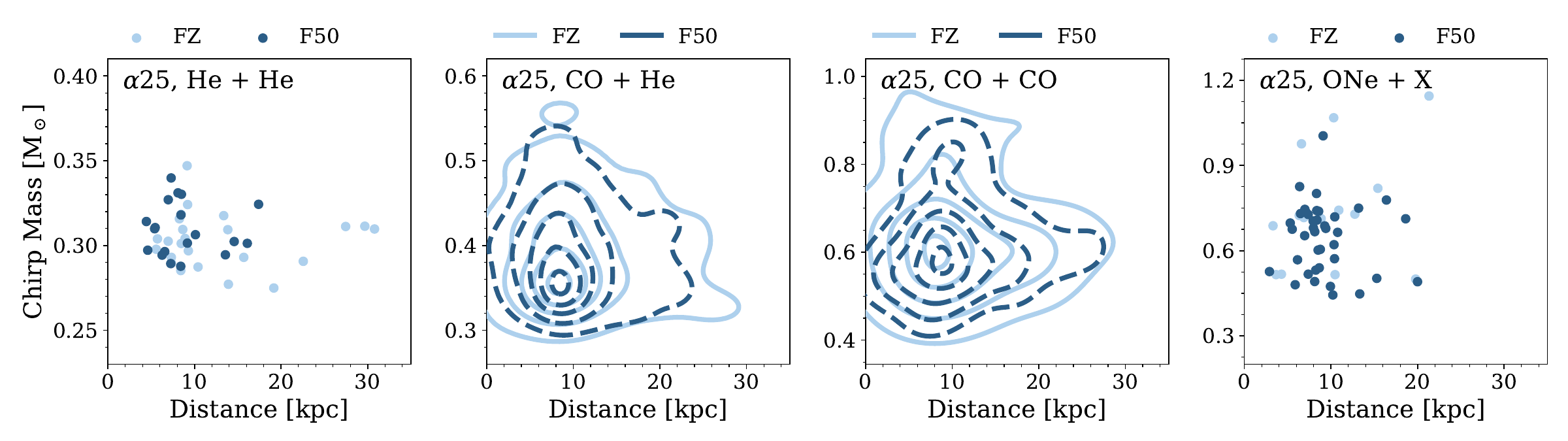}
	\includegraphics[width=0.8\textwidth]{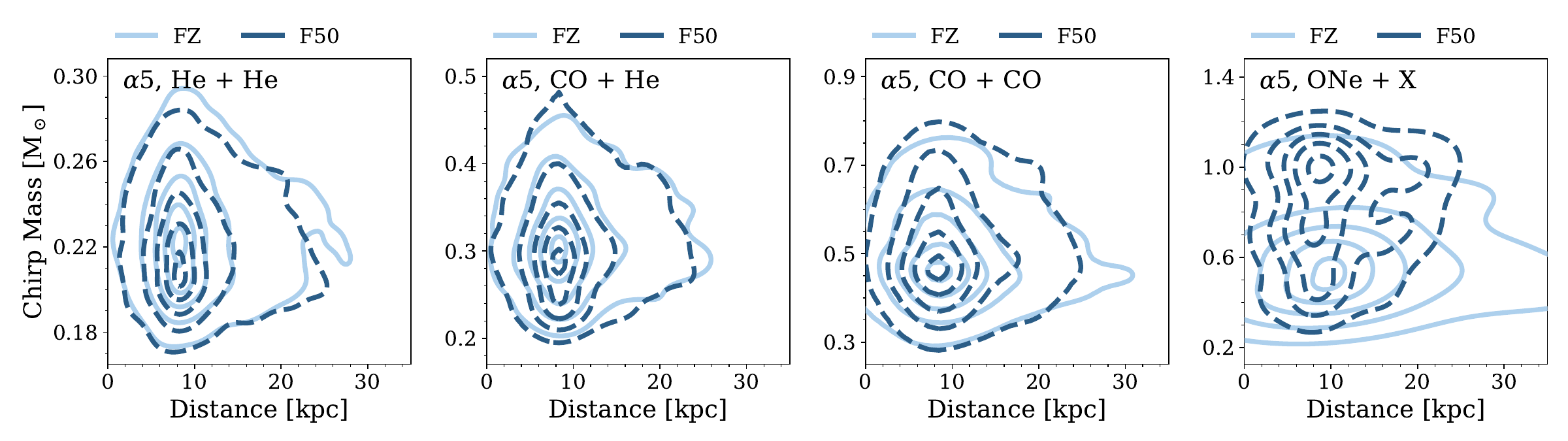}
	\includegraphics[width=0.8\textwidth]{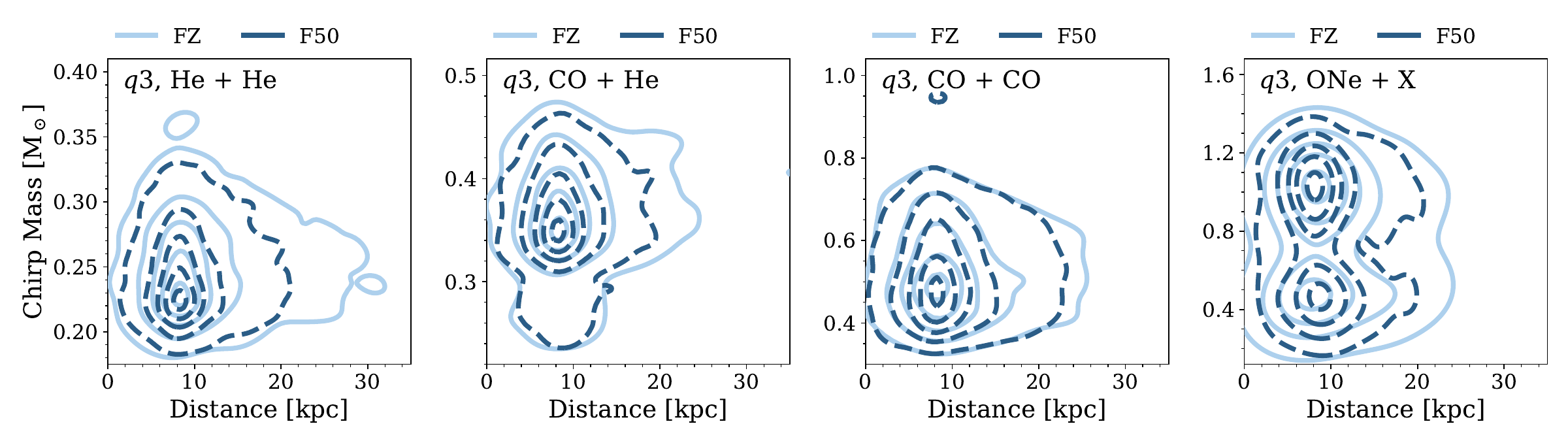}
    \caption{Chirp mass -- distance distributions for our other three binary evolution parameter variations. We plot the $\alpha25$ populations of He + He and ONe + X DWDs as scatter points since there are too few binaries to produce meaningful density distributions. Again, we have plotted each DWD type's population that exhibit observable GW frequency evolution and have $\rm{SNR}>7$. Contours show the 5$^{\rm{th}}$, 25$^{\rm{th}}$, 50$^{\rm{th}}$, 75$^{\rm{th}}$ and 95$^{\rm{th}}$ percentiles. Most populations remain unchanged between binary fraction models except for ONe + X DWDs in $\alpha5$, due to the number of He vs. CO vs. ONe companions between each model.}
    \label{fig:Mc_vs_dist_vars}
    \script{Mc_vs_dist.py}
\end{figure*}

Figure~\ref{fig:Mc_vs_dist_vars} shows the chirp mass vs. distance distribution for the $\alpha25$, $\alpha5$, and $q3$ parameter variations. For $\alpha25$ we plot scatter plots for He + He and ONe + X due to low statistics; there are too few binaries in this selected population to produce reliable kernel density estimates. The only population with a significant change between binary fraction models is for ONe + X DWDs, variation $\alpha5$. Here, we find a skew to lower chirp masses because a higher fraction of the binaries are ONe + He for FZ compared to F50: a Helium companion makes up 38\% of this selected population for FZ, compared with 12\% for F50. This leads to lower average chirp masses for the ONe + X distribution. Furthermore, $\sim6$\% of the F50 measurable systems are ONe + ONe binaries, versus 0\% for FZ. Because ONe WDs have larger masses, this also creates the high-mass concentration we find in the F50 contours.
\end{document}